%% file: JHEP-fluxesv2.tex
\documentclass[11pt,a4paper]{article}                           

\usepackage[bookmarksopen, bookmarksnumbered, bookmarksopenlevel=2]{hyperref}
\usepackage{tikz}
\usepackage{tikz-3dplot}
\usetikzlibrary{calc}
\usetikzlibrary{decorations} %
\usepackage[UKenglish]{babel}
\usepackage[toc,page]{appendix}
\usepackage{amsmath}
\usepackage{amssymb}
\usepackage{graphicx}
\usepackage{hhline}
\usepackage[bf]{caption}
\usepackage{cite}
\usepackage[vcentermath]{youngtab}
\usepackage{geometry}
\DeclareSymbolFont{bbold}{U}{bbold}{m}{n}
\DeclareSymbolFontAlphabet{\mathbbold}{bbold}

\usepackage{todonotes}

\usepackage{cancel}

\hypersetup{
    pdftitle={Fluxes in F-theory Compactifications on Genus-One Fibrations},
    pdfauthor={Ling Lin, Christoph Mayrhofer, Oskar Till, Timo Weigand},
    pdfsubject={F-theory compactifications}
}
\numberwithin{equation}{section}
\numberwithin{table}{section}

\newcommand{\bea}{\begin{eqnarray}}
\newcommand{\eea}{\end{eqnarray}}

\newcommand{\tw}{{\rm w}}

\newcommand{\executeiffilenewer}[3]{%
 \ifnum\pdfstrcmp{\pdffilemoddate{#1}}%
 {\pdffilemoddate{#2}}>0%
 {\immediate\write18{#3}}\fi%
}

\begin{document}

\baselineskip=14pt
\parskip 5pt plus 1pt

\vspace*{-1.5cm}
\begin{flushright}    
  {\small

  }
\end{flushright}

\vspace{2cm}
\begin{center}        
  {\LARGE Fluxes in F-theory Compactifications on Genus-One Fibrations}
\end{center}

\vspace{0.75cm}
\begin{center}        
Ling Lin$^1$, Christoph Mayrhofer$^2$, Oskar Till$^1$ and  Timo Weigand$^1$
\end{center}

\vspace{0.15cm}
\begin{center}        

\emph{$^1$Institut f\"ur Theoretische Physik, Ruprecht-Karls-Universit\"at, \\
             Philosophenweg 19, 69120
             Heidelberg, Germany\\
             }
             
             \emph{$^2$Arnold-Sommerfeld-Center, Ludwig-Maximilians-Universit\"at, \\
             Theresienstra{\ss}e 37, 80333
             M\"unchen, Germany\\
             }

\end{center}

\vspace{2cm}


\begin{abstract}
\noindent We initiate the construction of gauge fluxes in F-theory compactifications on genus-one fibrations which only have a multi-section as opposed to a section.
F-theory on such spaces gives rise to discrete gauge symmetries in the effective action.
We generalize the transversality conditions on gauge fluxes known for elliptic fibrations by taking into account the properties of the available multi-section.
We test these general conditions by constructing all vertical gauge fluxes in a bisection model with gauge group $SU(5) \times \mathbb Z_2$. The non-abelian anomalies are shown to vanish.
These flux solutions are dynamically related to fluxes on a fibration with gauge group $SU(5) \times U(1)$ by a conifold transition. 
Considerations of flux quantization reveal an arithmetic constraint on certain intersection numbers on the base which must necessarily be satisfied in a smooth geometry.
Combined with the proposed transversality conditions on the fluxes these conditions are shown to imply cancellation of the discrete $\mathbb Z_2$ gauge anomalies as required by general consistency considerations.

\end{abstract}

\thispagestyle{empty}
\clearpage
\tableofcontents
\thispagestyle{empty}


\newpage
\setcounter{page}{1}

\section{Introduction}

The portion of the F-theory \cite{Vafa:1996xn,Morrison:1996na,Morrison:1996pp}  landscape known to be populated by consistent vacua has considerably grown in the past year. This is partly due to the realization that F-theory compactifications do not necessarily require the existence of a section in order to be fully consistent \cite{Braun:2014oya}.
Indeed, the typical assertion that F-theory is defined in terms of an elliptic fibration can be  weakened in favor of the requirement that the underlying Calabi-Yau fourfold need only be genus-one fibered.
While it is by now a commonplace that F-theory is defined in terms of a torus whose complex structure geometrizes the axio-dilaton of $SL(2,\mathbb Z)$-invariant strongly coupled Type IIB theory (see e.g.\ \cite{Denef:2008wq,Weigand:2010wm} for reviews), the difference between elliptic and genus-one fibrations has been widely appreciated only recently: An elliptic fibration possesses a section which defines an embedding of the base into the Calabi-Yau fourfold, while a genus-one fibration only comes with a multi-section which realizes an embedding of a multi-cover of the base into the fourfold. 
Amongst the physical implications of the replacement of a section by an $n$-section is  the appearance of a discrete $\mathbb Z_n$ gauge group factor in the 4-dimensional effective action \cite{Morrison:2014era,Anderson:2014yva,Klevers:2014bqa,Garcia-Etxebarria:2014qua,Mayrhofer:2014haa,Mayrhofer:2014laa,Cvetic:2015moa}. 
Indeed, similarly to the zero-section of an elliptic fibration, the embedding multi-section of a genus-one fibration generates a massless $U(1)$ in the 3-dimensional effective action obtained by dimensional reduction of M-theory on the fourfold \cite{Anderson:2014yva,Mayrhofer:2014haa}. In an elliptic fibration, this Kaluza-Klein (KK) $U(1)$ becomes part of the 4-dimensional Lorentz group in the F-theory limit.
For $n$-section models, by contrast, a $\mathbb Z_n$ subgroup of the Kaluza-Klein $U(1)$ survives the F-theory limit as an extra, independent discrete gauge group factor (see \cite{Mayrhofer:2014haa,Mayrhofer:2014laa} for a detailed account of the origin of this extra symmetry). Apart from being interesting by itself, this mechanism is behind the realization of discrete gauge symmetries in particle physics applications \cite{Heckman:2010bq,Maharana:2012tu} of F-theory, the easiest being an R-parity \cite{Garcia-Etxebarria:2014qua,Mayrhofer:2014haa}.  Such discrete symmetries have been studied extensively in the weakly coupled Type II regime recently both from a general point of view \cite{BerasaluceGonzalez:2011wy,Camara:2011jg,BerasaluceGonzalez:2012zn,BerasaluceGonzalez:2012vb,Berasaluce-Gonzalez:2013bba,Berasaluce-Gonzalez:2013sna} and with an eye to phenomenological applications \cite{Ibanez:2012wg,Anastasopoulos:2012zu,Honecker:2013hda,Honecker:2015ela}.

The  geometric structure underlying  discrete gauge groups in F-theory is in fact even richer. Associated to a genus-one fibration with an $n$-section is a set of $n$ inequivalent fibrations with the same Jacobian. These  are counted by the Tate-Shafarevich group associated with this Jacobian \cite{Braun:2014oya,deBoer:2001px}. Each of the $n$ fibrations gives rise to a different M-theory background, which all map to the same F-theory effective action in 4 dimensions \cite{Morrison:2014era,Mayrhofer:2014haa,Mayrhofer:2014laa,Cvetic:2015moa}. Of these the Jacobian fibration takes a distinguished role in that it is the only one where a discrete symmetry is realized already in the M-theory effective action \cite{Mayrhofer:2014laa}. It is also, by definition, the only one which is elliptically fibered. The origin of the discrete gauge group is here very different and rooted in the appearance of torsional homology \cite{Mayrhofer:2014laa} as expected from the general arguments of \cite{Camara:2011jg,Grimm:2011tb}.  The remaining $n-1$ genus-one fibrations are related to an underlying elliptic fibration by a topological transition which describes the Higgsing with a 3-dimensional field of non-trivial KK charge. The resulting background can equivalently be described in terms of a fluxed $S^1$ reduction \cite{Anderson:2014yva,Mayrhofer:2014haa,Mayrhofer:2014laa}. 
This leads to a beautiful picture unifying aspects of arithmetic geometry, torsional cohomology and non-standard KK reductions in a physical framework which, last but not least, has interesting phenomenological properties. 


A natural next step in the investigation of these new F-theory backgrounds is the inclusion of 4-form background fluxes. 
In F/M-theory both the `closed string' fluxes and the gauge fluxes are known to enjoy a unified description in terms of the $G_4$-fluxes of 11-dimensional supergravity \cite{Becker:1996gj}. 
Our focus here will be on the gauge sector. The explicit construction of such $G_4$ gauge fluxes in globally defined F-theory compactifications on elliptic fibrations has been studied in great detail recently, including the works \cite{Collinucci:2010gz,Grimm:2011tb,Braun:2011zm,Marsano:2011hv,Krause:2011xj,Grimm:2011fx,Krause:2012yh,Collinucci:2012as,Tatar:2012tm,Kuntzler:2012bu,Cvetic:2012xn,Mayrhofer:2013ara,Cvetic:2013uta,Borchmann:2013hta,Cvetic:2013jta,Braun:2014pva,Bies:2014sra,Bizet:2014uua,Braun:2014xka,Cvetic:2015txa,Watari:2015ysa}.
First steps in extending these results to F-theory compactifications without a section have already been undertaken in \cite{Mayrhofer:2014haa}. In this article we will systematize the construction of gauge fluxes in multi-section fibrations. 

The first question  is how to generalize the consistency conditions governing the correct uplift of $G_4$-fluxes from M-theory to purely internal fluxes in F-theory. 
As we will review in section \ref{sub_transversality}, in elliptic fibrations these conditions are a formalization (see e.g.\ \cite{Donagi:2008ca,Braun:2011zm,Marsano:2011hv,Krause:2011xj,Grimm:2011fx}) of the transversality conditions which go back in essence to the early work of \cite{Dasgupta:1999ss}. These admit an equivalent characterization in the language of 3-dimensional supergravity \cite{Grimm:2010ks,Grimm:2011fx,Cvetic:2012xn}.
We will argue how to generalize the transversality conditions by replacing the embedding zero-section of an elliptic fibration by the embedding of the multi-cover of the base provided by the multi-section. 
This is based on the aforementioned identification of a KK $U(1)$ associated with the multi-section which parallels the procedure in elliptic fibrations. In general the class of the embedding multi-section must be corrected such as to suitably normalize the KK $U(1)$ in comparison with the Cartan $U(1)$ of potential non-abelian gauge groups. 
We will then subject this transversality condition to a number of non-trivial tests by explicitly constructing, in section \ref{sec_vertfluxZ2}, all vertical flux solutions for a bisection fibration with F-theory gauge group $SU(5) \times \mathbb Z_2$, whose geometry is reviewed from \cite{Mayrhofer:2014haa} (see also \cite{Garcia-Etxebarria:2014qua}) in section \ref{sec_SU5Z2}. 
We will follow our previous approach \cite{Krause:2012yh} and assume that the fibration is defined over a generic base space ${\cal B}$. This way we focus on those flux solutions which are guaranteed to exist for any such base. In addition we generalize a specific horizontal gauge flux constructed already in \cite{Mayrhofer:2014haa}. 
The vertical flux solutions are shown to be related to the cohomology classes of the matter surfaces in section \ref{sec:z2-matter_surfaces}.
We explicitly compute the chiral spectrum in section \ref{sec_Chiralities} and, as a first consistency check, confirm that the transversality conditions on the gauge fluxes imply the cancellation of the non-abelian gauge anomalies.

The bisection geometry is related, via a conifold transition, to the elliptic fibration with gauge group $U(1)$ introduced in \cite{Morrison:2012ei}.
On general grounds, the gauge fluxes on both sides of the conifold transition must be related in such a way that topological invariants such as the total induced D3-brane charge and the chiralities with respect to unbroken gauge groups remain invariant. 
In section \ref{sec:u1_geometry} we construct all vertical gauge fluxes for the $SU(5) \times U(1)$ model, employing the same methods as on the bisection side. These are then shown to dynamically match with the fluxes of the bisection fibration upon performing a conifold transition between both models. This is another consistency check of the flux construction. Similar matchings had been demonstrated before for the conifold transition relating the $SU(5) \times U(1)$ restricted Tate model to a generic $SU(5)$ Tate model \cite{Braun:2011zm,Krause:2012yh}.

Finally, in section \ref{sec_Comparison}, we address two subtle and related issues, the quantization condition \cite{Witten:1996md} for gauge fluxes and the cancellation of discrete gauge anomalies. 
For the explicit $SU(5) \times \mathbb Z_2$ and $SU(5) \times U(1)$ fibrations under consideration, we derive a certain arithmetic constraint on the intersection numbers of the base divisor classes defining the fibration which is necessary and sufficient to obtain an integral chiral spectrum. These constraints arise by requiring that $\frac{1}{2} c_2(M_4)$ must integrate to an integer over every matter surface. We conjecture that this constraint is automatically satisfied for any \emph{smooth} fibration of the considered types. Assuming this condition to hold we show that the discrete $\mathbb Z_2$ gauge anomalies are automatically cancelled for any flux which satisfies our modified transversality condition. This is indeed required \cite{Ibanez:1992ji,BerasaluceGonzalez:2011wy} because for the type of genus-one fibrations considered here, the discrete symmetry is non-perturbatively exact \cite{Martucci:2015dxa,Martucci:2015oaa}.
This is the final non-trivial check for our construction.

\section{Fluxes on a $\mathbb P_{112}[4]$-fibration with a bisection }\label{sec:bisection_geometry}

In this section we propose a generalization of the well-known transversality conditions on $G_4$-fluxes to fibrations with a multi-section only. We will test this general proposal by constructing  all vertical fluxes
plus a certain type of horizontal gauge flux on  a $\mathbb P_{112}[4]$-fibration $X_4$ with a bisection over a generic base ${\cal B}$.
In addition we will engineer  an extra $SU(5)$ singularity.
Apart from specifying the cohomology groups of the gauge fluxes as elements of $H^{2,2}(X_4)$, we will identify explicit algebraic four-cycles whose associated rational equivalence classes parametrise the gauge data in the sense described in \cite{Bies:2014sra}. 
For the vertical fluxes these are precisely the matter surfaces corresponding to the $SU(5)$ charged matter states. We will compute the chiral spectrum in full generality and show that the fluxes induce no $SU(5)$ anomaly.

\subsection{The transversality condition for multi-section fibrations} \label{sub_transversality}

Our proposal for a modified notion of transversality for gauge fluxes in F-theory compactifications without section
 will apply to a general genus-one fibration $X_4$ with projection
\bea
\pi: X_4 \rightarrow {\cal B}
\eea
over a generic 3-dimensional base space ${\cal B}$.  A genus-one fibration which is not an elliptic fibration can in general have several independent multi-sections \cite{Braun:2014oya}, but no holomorphic or rational section.
Since the transversality conditions on elliptic fibrations make use of the notion of a zero-section, we first need to review the origin of these conditions for conventional elliptic fibrations and then extend them to situations without sections.


The flux transversality conditions derive from the behaviour of the 4-form flux $G_4$ in the standard F-theory limit of M-theory compactifications on fourfolds (see e.g.~\cite{Denef:2008wq} for a review). 
By compactifying 11-dimensional supergravity on a $T^2$-fibered fourfold $M_4$ a 3-dimensional $\mathcal N = 2$ field theory is obtained.
 In the F-theory limit  the 3-dimensional compactification lifts to a 4-dimensional theory. The limit amounts to sending the fiber volume Vol$(T^2)$ to zero. Denoting the radii of the two one-cycles of the torus by $R_A$ and $R_B$, the limit is taken in two steps. First, the $A$-cycle is identified with the M-theory circle, and the limit $R_A \rightarrow 0$ is the weakly coupled type IIA limit of M-theory. The second step is a T-duality transformation along the $B$-cycle, which gives type IIB on a circle of radius $\tilde{R}_B = \frac{l_s^2}{R_B}$. In the compactification limit $R_B \rightarrow 0$ the dual circle decompactifies and one ends up with a (generically strongly coupled) type IIB theory in four dimensions. Importantly, one of the four large dimensions  has its origin in one of the fiber directions of the fourfold. One immediate consequence is that care must be taken when introducing fluxes \cite{Sethi:1996es,Becker:1996gj}. 4-dimensional Lorentz invariance forbids fluxes with non-trivial VEV along the circle along which the T-dualization is performed \cite{Dasgupta:1999ss}.
 \noindent More precisely, the $G_4$ flux must have \textit{one leg in the fiber} to meet this requirement. Indeed, in \cite{Dasgupta:1999ss} it was shown that a flux with zero or two legs along the fiber maps to the self-dual 5-form flux $F_5$ in type IIB string theory. In this case the vacuum expectation value extends along the non-compact directions and breaks Lorentz invariance. The remaining possibility is a flux with one leg in the fiber. These solutions do not lie completely in the base, nor do they fill the two fiber directions. 

This transversality condition \cite{Dasgupta:1999ss} is usually expressed in slightly more formal terms as follows: Let us first consider the standard case of an \emph{elliptically} fibered fourfold
\bea
\pi_M: M_4 \rightarrow {\cal B}
\eea
with projection $\pi_M$. By definition, an elliptic fibration has a zero section $\sigma^{(0)}: {\cal B} \rightarrow M_4$ which defines an embedding of the base ${\cal B}$ as a divisor $\sigma^{(0)}({\cal B})$ into $M_4$,
\bea
\iota_\sigma: \sigma^{(0)}({\cal B}) \hookrightarrow M_4.
\eea
In the presence of several independent sections, i.e.\ for an elliptic fibration with a Mordell-Weil group of non-zero rank, the choice of zero-section is not unique \cite{Grimm:2013oga,Grimm:2015zea}, but the different choices all asymptote to the same effective theory in the F-theory limit. 

Let us therefore assume that we have singled out one particular section as our zero-section and denote by $Z$ its homology class. For simplicity we assume the zero-section to be holomorphic, but this is not necessary \cite{Grimm:2013oga,Cvetic:2013nia}. From the perspective of the 3-dimensional M-theory effective action, $Z$ generates a $U(1)$ gauge group which is to be identified with the Kaluza-Klein $U(1)$ obtained by reducing the 4-dimensional F-theory compactification along a circle $S^1$ (see \cite{Grimm:2010ks} for a recent discussion in the language of 3-dimensional supergravity). In the effective action light charged matter states arise from M2-branes wrapping suitable fibral curves \cite{Witten:1996qb,Katz:1996xe,Intriligator:1997pq,LieF}. This includes both the non-Cartan vector bosons and related matter states and extra charged localised matter. 
More precisely, each component field $\Psi(x,z)$  of an ${\cal N}=1$ multiplet of the 4-dimensional F-theory action decomposes, upon circle reduction to  three dimensions, to a zero mode plus a full tower of Kaluza-Klein excitations  $\Psi(x,z) = \sum_{n \in \mathbb Z} \psi_n(x) e^{inz}$. Here $x$ denotes external coordinates in the 3-dimensional M-theory vacuum and $z$ is the KK-circle coordinate.
The higher KK states have KK $U(1)$ charge $n = \int_{C_n} Z$, where $C_n$ is the fibral curve wrapped by the M2-brane associated with state $\psi_n(x)$.
Since $Z$ is a section, it has intersection number $+1$ with a generic non-degenerate fiber. 
This is still true for split fibers in higher codimension, but not all components of the fiber will intersect $Z$. 
 Thus, the zero mode $\psi_0$ is due to M2-branes wrapping a  fibral curve $C_0$ with vanishing intersection with the zero-section $Z$. The KK partner of KK charge $n$ is then created by an M2-brane wrapping in addition the full torus elliptic fiber ${\mathfrak f}$ $n$-times such that its associated fibral curve can be written as $C_n = C_0 + n \, {\mathfrak f}$.

At the cohomological level,  the transversality condition of \cite{Dasgupta:1999ss} on gauge fluxes is that  (e.g.\ \cite{Donagi:2008ca,Braun:2011zm,Marsano:2011hv,Krause:2011xj,Grimm:2011fx})
\begin{align}\label{eq:transv_cond_standard_1}
\int_{M_4} G_4 \wedge Z \wedge  \pi_M^{-1}D_a &\stackrel{!}{=} 0,\\
\label{eq:transv_cond_standard_2}
\int_{M_4} G_4 \wedge  \pi^{-1}_M D_a \wedge  \pi^{-1}_M D_b &\stackrel{!}{=}  0 \,
\end{align}
for $D_{a,b}$ any divisor classes on the base.\footnote{For ease of notation we will oftentimes be laid-back and omit the explicit pull-back symbol.} 
 The first condition \eqref{eq:transv_cond_standard_1}  guarantees that $G_4$ does not lie completely in the base because it requires that 
  \bea \label{intZa}
\int_{M_4} G_4 \wedge Z \wedge \pi^{-1}_MD_a = \int_{\sigma^{(0)}(\mathcal{B})} \iota_\sigma^* (G_4 \wedge  \pi^{-1}_MD_a)   \stackrel{!}{=}  0,
\eea
 i.e.\ the net flux  through any four-cycle $D_a$ on the base vanishes. 
The second condition \eqref{eq:transv_cond_standard_2} expresses that the solution cannot have two (real) legs along the fiber. 
This condition can be rephrased as the constraint that the chiral index of all KK partners equals that of the zero mode. Indeed 
 the intersection $\pi_M^{-1}D_a \cap \pi_M^{-1}D_b$ is a four-cycle extending along the full fiber over a curve $D_a \cap D_b$ in the base and $\int_{\pi_M^{-1}D_a \cap \pi_M^{-1}D_b} G_4$ computes the chiral index of states associated with M2-branes wrapping the full fiber over $D_a \cap D_b$. If the integral over any four-cycle of this type vanishes, this guarantees in particular that the multiplicities of the fields $\psi_n$ are the same for all $n$.   This is the field theoretic way of stating the requirement of Lorentz invariance. A discussion along these lines can also be found e.g.\ in \cite{Cvetic:2012xn,Cvetic:2013uta}.

%
In models with non-abelian gauge symmetries the Cartan generators correspond to the exceptional divisor classes $E_i$ from the resolution of the singularity. In order to leave the non-abelian gauge group unbroken in the F-theory limit, we must in addition demand that
\begin{equation} \label{G4Ei}
\int_{M_4} G_4 \wedge E_i \wedge \pi_M^{-1}D_a \stackrel{!}{=} 0 \, .
\end{equation}
Indeed, M2-branes wrapping combinations of the rational fibers $\mathbb P^1_i$ of the resolution divisors $E_i$ give rise to non-abelian massless vector bosons in the F-theory limit \cite{Witten:1996qb}. The condition (\ref{G4Ei}) guarantees that the flux induces no chiral index for the associated gauginos.
If one of these conditions fails, the F-theory gauge group will be broken to the commutant of the associated Cartan generator. This is utilized in models with hypercharge GUT breaking \cite{Beasley:2008kw, Donagi:2008kj}, but our focus will be on non-Cartan fluxes in this paper. Note also that a holomorphic zero-section intersects precisely the affine node of the Kodaira fiber over a divisor with non-abelian gauge group. Thus the condition \eqref{eq:transv_cond_standard_1} is fulfilled by all the Cartan fluxes $E_i \wedge \pi_M^{-1}F$ for any class $F$ pulled back from the base. Therefore the special case of \eqref{eq:transv_cond_standard_1} for $G_4 = E_i \wedge \pi_M^{-1} F $ is the condition that the KK $U(1)$ is chosen `orthogonal' to the non-abelian gauge group.

We are now in a position to generalize these criteria to F-theory compactifications on non-elliptic genus-one fibrations. 
 As stressed above in such geometries no (holomorphic or rational) section exists, but only one or several multi-sections.
An $n$-section  is a multi-valued map assigning to each point in the base locally $n$-points in the fiber which are globally exchanged by monodromies. This defines an $n$-fold branched cover $\mu_n({\cal B})$ of the base ${\cal B}$ inside $X_4$ together with an embedding
\bea
\iota_\mu: \mu_n({\cal B}) \hookrightarrow X_4.
\eea
Let us denote the homology class of the $n$-section as $N$.
For the purpose of relating the M-theory reduction to F-theory it is necessary to specify a notion of KK $U(1)$. As pointed out several times by now, it is still true that a multi-section defines such a KK $U(1)$ similarly to the case of an elliptic fibration \cite{Anderson:2014yva,Garcia-Etxebarria:2014qua,Mayrhofer:2014haa,Mayrhofer:2014laa} because it is possible to expand the M-theory 3-form $C_3$ as $C_3 = A_{KK} \wedge N + \ldots$. 
We therefore need to choose an $n$-section as the substitute for the zero-section and define transversality with respect to the associated KK frame.

In terms of this embedding multi-section we then have
\bea
\int_{X_4} G_4 \wedge N \wedge \pi^{-1} D_a = \int_{ \mu_n({\cal B}) } \iota_n^*(G_4 \wedge \pi^{-1} D_a).
\eea
Therefore the analogue of \eqref{eq:transv_cond_standard_1} is
\bea \label{eq:nsection_one_leg}
\int_{X_4} G_4 \wedge N \wedge \pi^{-1} D_a \stackrel{!}{=}0,
\eea
which guarantees that the net flux vanishes through every base four-cycle.
Second, since the multi-section still defines the notion of a KK $U(1)$, the condition that all elements of the KK tower should have the same chiral index implies that the analogue of \eqref{eq:transv_cond_standard_2} must still hold.

 In general the $n$-section intersects more than one of the irreducible curves in the Kodaira fiber over a divisor with non-abelian gauge group. This implies that the Cartan fluxes do not satisfy \eqref{eq:nsection_one_leg} in general. However, as we will see in the explicit example below, one may construct a divisor class
\begin{equation}\label{eq:nsection_hat}
\hat{N} = N + \sum_i a_i E_i
\end{equation}
and choose the coefficients $a_i$ such that the modified condition
\begin{equation}\label{eq:mod_transv_condition}
\int_{X_4} G_4 \wedge \hat{N} \wedge \pi^{-1} D_a = 0
\end{equation} 
is satisfied for  Cartan fluxes $G_4 = E_i \wedge \pi^{-1} F$ for any class $F \in H^{1,1}({\cal B})$. This is a linear system of equations for the coefficients $a_i$ with a unique solution. This is because the Cartan matrices for the simple Lie algebras, which appear as intersection numbers in \eqref{eq:mod_transv_condition} through
\bea
\int_{X_4} E_i \wedge E_j \wedge \pi^{-1} \omega_2 = - C_{ij} \int_\Theta \, \omega_2, \qquad \quad \omega_2 \in H^2({\cal B}),
\eea
are invertible. Here $\Theta$ is the divisor supporting non-abelian gauge symmetry in the base. The choice of $\hat N$ amounts to a redefinition of the KK $U(1)$ symmetry such that it does not mix with the Cartan $U(1)$ generators $E_i$ associated with the resolution divisors of the non-abelian singularity. A similar redefinition has been discussed in a different context in \cite{Grimm:2015zea}.



To summarize our proposal for the transversality condition in a genus-one fibration: Fix a multi-section class $N$ defining the embedding of an $n$-fold cover of the base into $X_4$ and determine $\hat N = N + \sum_i a_i E_i$ such that 
\bea \label{orthogonality-condition}
\int_{X_4} E_i \wedge \hat N \wedge \pi^{-1} D_a \wedge  \pi^{-1} D_b = 0   \qquad \quad \forall D_a, D_b.
\eea
This $\hat N$ defines the appropriate KK $U(1)$ for the reduction of the 4-dimensional F-theory vacuum to three dimensions which does not mix with the Cartan generators $E_i$. The transversality conditions on the fluxes are then
\bea
&& \int_{X_4} G_4 \wedge \hat N \wedge \pi^{-1} D_a \stackrel{!}{=}0, \label{transversality-nsection1}  \\
&& \int_{X_4} G_4 \wedge \pi^{-1} D_a  \wedge  \pi^{-1} D_b \stackrel{!}{=}0. \label{transversality-nsection2}  
\eea
If the gauge fluxes are not to break any of the non-abelian gauge symmetries, we demand in addition 
\bea \label{Eicond}
 \int_{X_4} G_4 \wedge E_i \wedge \pi^{-1} D_a \stackrel{!}{=}0. 
\eea
Note in particular that for fluxes satisfying this latter constraint for all $E_i$, the transversality condition (\ref{transversality-nsection1}) reduces to the same constraint with $\hat N$ replaced by $N$.
This simplifies the calculations, but obscures the fact that $\hat{N}$ is the divisor class identified with the Kaluza-Klein $U(1)$.

\subsection[A genus-one fibration with gauge group {$SU(5) \times \mathbb Z_2$}]{A genus-one fibration with gauge group {\boldmath{$SU(5) \times \mathbb Z_2$}} } \label{sec_SU5Z2}


We will now briefly review the bisection $\mathbb P_{112}[4]$-fibration with gauge group $SU(5) \times \mathbb Z_2$ which will serve as our laboratory to illustrate and test the ideas presented in the previous section.
We will stick to the notation of \cite{Mayrhofer:2014haa}, where this specific geometry was discussed in detail (see also \cite{Garcia-Etxebarria:2014qua}).
The fourfold $X_4$ is given by the hypersurface equation
\begin{equation}\label{eq:hypersurface-su5xZ2-model}
\begin{aligned}
P_{\mathbb Z_2}^{SU(5)}=&\,e_1 e_2 w^2 + b_{0,2} u^2 w e_0^2 e_1^2 e_2 e_4 + b_1 u v w + b_2 v^2 w e_2 e_3^2 e_4\\
& + c_{0,4} u^4 e_0^4 e_1^3 e_2 e_4^2 + c_{1,2} u^3 v e_0^2 e_1 e_4 + c_{2,1} u^2 v^2 e_0 e_3 e_4 + c_{3,1} u v^3 e_0 e_2 e_3^3 e_4^2 + c_{4,1} v^4 e_0 e_2^2 e_3^5 e_4^3
\end{aligned}
\end{equation}
in the toric ambient space specified in table \ref{tab:scalings_z2} of appendix \ref{app_scaling1}. It is a genus-one fibration \cite{Braun:2014oya,Anderson:2014yva,Morrison:2014era,Klevers:2014bqa,Garcia-Etxebarria:2014qua,Mayrhofer:2014haa,Mayrhofer:2014laa} over a general base $\mathcal B$.
The fiber coordinates $[u : v : w]$ are homogeneous coordinates of $\mathbb P_{112}$. An $SU(5)$ singularity sits in the fiber over the divisor $\Theta : \{\theta=0\}$ in $\mathcal B$. The hypersurface equation is the proper transform under the resolution of this singularity, with blow-up coordinates $e_i$, $i=1, \ldots,4$, and with  $\pi^*\theta=e_0\cdot e_1 \cdot \ldots \cdot e_4$. The Calabi-Yau hypersurface comes with the choice of a line bundle on ${\cal B}$ with first Chern class $[b_2]$. Given this line bundle on $\mathcal B$ the coefficients $b_i$ and $c_j$ transform as sections of the bundles displayed in table \ref{tab:coeff1}, where $\bar{\mathcal K}$ is the anti-canonical bundle on the base. 

\begin{table}[ht]
\centering 
\begin{tabular}{c|c|c|c|c|c|c|c}
 $b_{0,2}$ & $b_{1}$ & $b_{2}$ & $c_{0,4}$ & $c_{1,2}$ & $c_{2,1}$&$c_{3,1}$ & $c_{4,1}$ \\
\hline \rule{0pt}{.4cm}
 $2\bar{\mathcal K} - b_2 - 2 \Theta$& $\bar {\cal K}$ &  $b_2$ &   $4 \bar{\mathcal K} - 2 b_2 - 4\Theta$ &   $3\bar {\cal K} - b_2 - 2\Theta$ &    $ 2\bar {\cal K} -\Theta$ &   $\bar {\cal K} + b_2 - \Theta$ &   $2 b_2 - \Theta$\\
\end{tabular}
\caption{Classes of the coefficients entering \eqref{eq:hypersurface-su5xZ2-model}.}\label{tab:coeff1}
\end{table}

\noindent The smooth geometry is constructed via a top \cite{Candelas:1996su,Bouchard:2003bu}, denoted $\tau_{4,3}$ in \cite{Braun:2013nqa}, and the exceptional divisors are $E_i: \{e_i = 0\}$, $i=1, \ldots,4$.
Furthermore $E_0 = \pi^*\Theta - \sum_i E_i$. 
The Stanley-Reisner ideal for our choice of resolution phase is generated by
\begin{equation}\label{eq:SRi_z2}
\textmd{SR-i}:\quad \{v\,e_0,\, v\,e_1,\, v\,e_2,\, w\,e_0,\, w\,e_4,\, u\,e_3,\, e_0\,e_3,\, e_1\,e_3,\, u\,e_2,\, e_1\,e_4,\, v\,w\,u\}\, .
\end{equation}
The intersection of the ambient divisor $U: \{u = 0\}$ with the hypersurface gives a representative of the homology class of the bisection, which intersects each generic fiber in two points exchanged globally by a monodromy.
From our previous discussion we would like to associate with $U$ the notion of a KK $U(1)$ in the 3-dimensional M-theory compactification on $X_4$. It is here that the shift (\ref{eq:nsection_hat}) becomes important because the bisection locally intersects both $E_0$ and $E_1$ in one point in the fiber.
The (up to normalization) unique solution to the constraints (\ref{orthogonality-condition}) is given by  
\bea
\hat U  =U +  \frac{1}{5}(4 E_1 + 3 E_2 + 2 E_3 + E_4).
\eea
If we fix  the (a priori arbitrary) overall normalization to achieve integer intersections with all fibral curves by defining
\bea \label{twZ2def}
\tw_{\mathbb Z_2} = 5 \,  \hat U,
\eea
then the intersection numbers of $\tw_{\mathbb Z_2}$ with the irreducible split fiber components consistently assign $\mathbb Z_2$ charges to the corresponding states modulo 2 in the F-theory limit.
Indeed, a $\mathbb Z_2$ subgroup of the KK $U(1)$, normalised as in (\ref{twZ2def}), survives in the F-theory limit as an independent discrete gauge group  -- a full explanation can be found in \cite{Mayrhofer:2014haa,Mayrhofer:2014laa} (see also \cite{Morrison:2014era,Anderson:2014yva,Klevers:2014bqa,Garcia-Etxebarria:2014qua}).\footnote{Apart from an extra shift in terms of base divisors this agrees with the $\mathbb Z_2$ generator as presented in  \cite{Klevers:2014bqa,Garcia-Etxebarria:2014qua,Mayrhofer:2014haa}). This shift does not change the notion of fibral curves and is therefore not of importance for us.} 
\begin{figure}
\vspace{-1cm}
\centering \def\svgwidth{350pt} 
\hspace{1cm}
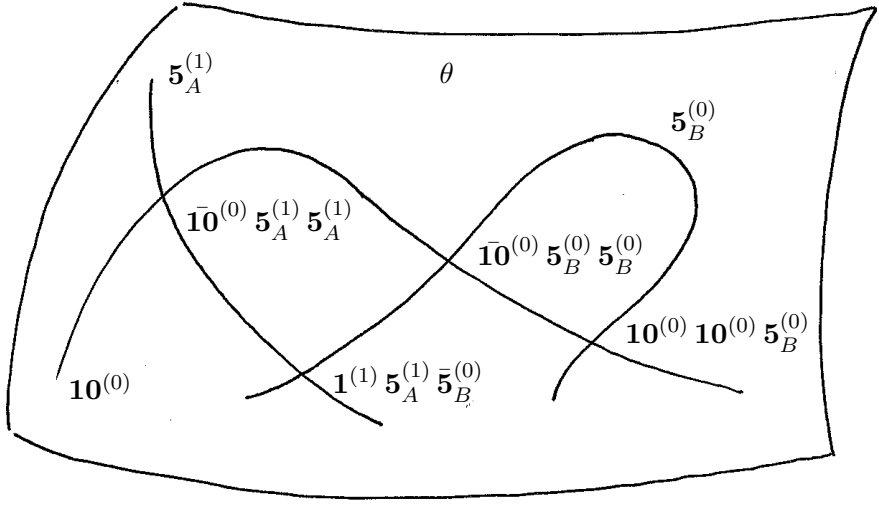 
\vspace{0cm}
\caption{The matter curves on the $SU(5)$ divisor $\{\theta = 0\}$ and the Yukawa couplings involving the $SU(5)$ charged matter in codimension three.}\label{fig:matter_curves_poly4}
\end{figure}
\noindent The discriminant of the hypersurface equation takes the form
\begin{equation} \label{discrZ2}
\Delta \sim \theta^5[b_1^4(b_1^2 c_{0,4} - b_0 b_1 c_{1,2} + c_{1,2}^2)(b_2^2 c_{2,1} - b_1 b_2 c_{3,1} + b_1^4 c_{4,1}) + \mathcal{O}(\theta)],
\end{equation}
which indicates  three matter curves on the $SU(5)$ divisor $\Theta$. Away from $\Theta$ there is one more matter locus \cite{Braun:2014oya}, describable as an ideal which defines an irreducible curve on ${\cal B}$ \cite{Mayrhofer:2014haa}.  This  complicated codimension-two locus $C_2$ over which the fiber is of type $I_2$ hosts singlet states that carry $\mathbb Z_2$ charge. These states originate from singly charged states in the $SU(5) \times U(1)$ model related to this geometry by a conifold transition. The matter spectrum and the associated $\mathbb Z_2$ charges are summarized in table \ref{table_matterZ2}.
\begin{table}[ht]
\begin{center}
\begin{tabular}{ l c r } 
locus in base & irrep $SU(5)$ & $\mathbb Z_2$ charge \\ 
\hline \rule{0pt}{.35cm}
$\theta \cap b_1$ & ${\bf{10}},{\bf{{\bar{10}}}}$  & [0] \\ \hphantom{\rule{0pt}{.35cm}}
$\theta \cap \{b_1^2 c_{0,4} - b_0 b_1 c_{1,2} + c_{1,2}^2\}$  & ${\bf{5}}^A,{\bf{{\bar{5}}}}^A$ & [1] \\ \hphantom{\rule{0pt}{.35cm}}
$\theta \cap \{b_2^2 c_{2,1} - b_1 b_2 c_{3,1} + b_1^4 c_{4,1}\}$ & ${\bf{5}}^B,{\bf{{\bar{5}}}}^B$ & [0] \\ \hphantom{\rule{0pt}{.35cm}}
$C_2$ & \bf{1} & [1] \\ 
\end{tabular}
\end{center}
\caption{Matter spectrum in the $SU(5) \times \mathbb Z_2$ model. \label{table_matterZ2}}
\end{table}
The intersection structure of the matter curves along the $SU(5)$ divisor $\Theta$ is shown in figure \ref{fig:matter_curves_poly4}, which we reproduce from \cite{Mayrhofer:2014haa} for convenience.
The indicated Yukawa couplings are  all consistent with the $\mathbb Z_2$ charges.



 \subsection[Horizontal and all vertical fluxes in an $SU(5) \times \mathbb Z_2$ model]{Horizontal and all vertical fluxes in the {\boldmath{$SU(5) \times \mathbb Z_2$}} model}  \label{sec_vertfluxZ2}

We can finally analyze the proposed transversality conditions for consistent $G_4$-fluxes,
\begin{equation}\label{eq:bisection_transv_cond}
\begin{aligned}
\int_{X_4} G_4 \wedge \hat{U} \wedge \pi^{-1}D_a &= 0,\\
\int_{X_4} G_4 \wedge D_a \wedge  \pi^{-1}D_b &= 0, \\
\int_{X_4} G_4 \wedge E_i \wedge  \pi^{-1}D_a &= 0 \,,
\end{aligned}
\end{equation}
in the $SU(5) \times \mathbb Z_2$ geometry, where the last condition only applies if we require the flux solution $G_4$ to leave a full $SU(5)$ gauge group unbroken in the F-theory limit. As noted before, this eliminates the correction terms in $\hat{U}$ and reduces the system to the usual transversality conditions with respect to the unshifted bisection $U$. 
Before explicitly solving these constraints, let us note that consistent $G_4$-fluxes are in addition subject to the quantization condition \cite{Witten:1996md,Collinucci:2010gz,Collinucci:2012as}
\bea \label{quantisation1}
G_4 + \frac{1}{2} c_2(X_4) \in H^4(X_4, \mathbb Z) \cap H^{2,2}(X_4).
\eea
We will come back to the subtle question of how to properly quantize the fluxes in section \ref{sec_quantisation} and for now focus on solving (\ref{eq:bisection_transv_cond}) without fixing the overall normalization.

Our first such flux is an example of a horizontal gauge flux which generalizes the horizontal $G_4$ flux constructed in \cite{Mayrhofer:2014haa} for the bisection model without further non-abelian gauge enhancement.
 The flux is associated with a special algebraic four-cycle which appears on the sublocus in complex structure moduli space where $c_{4} = \rho \,  \tau$. This is modeled after a similar construction in the context of a Tate model \cite{Braun:2011zm}. In the presence of an $SU(5)$ singularity the same type of fluxes exists, \emph{mutatis mutandis}, on the sublocus in moduli space where  $c_{4,1} = \rho \,  \tau$. 
In this case the two algebraic four-cycles described as the complete intersections
\bea
\sigma_0 &=& (u, w, \rho), \\
\sigma_1 &=& (u, w e_1 + b_2 v^2 e_3^2 e_4, \rho)
\eea
in the ambient space $X_5$ of $X_4$ automatically lie on $X_4$.  This notation indicates that the four-cycles should be thought of as the algebraic varieties associated with 
the ideal  generated by the polynomials in brackets.

The two four-cycles each define one of the two intersection points of the bisection $U$ with the fiber, fibered over the divisor $P : \{\rho = 0 \} $ in the base. The dual cohomology classes $[\sigma_0]$ and $[\sigma_1]$ are candidates for a flux. To obtain a well-defined flux we add an ansatz of correction terms $ \sum a_i D_i \wedge P$ where $D_i$ runs over a basis of divisors in the fourfold. Solving for the coefficients $a_i$ yields the flux solutions 
\bea \label{eq:z2_flux_sigma1}
G_4(P,\sigma_0) &=& 5 [\sigma_0] +  \frac{1}{2}(-5U + (4E_1 + 3E_2 + 2E_3+ E_4)- 2\theta )\wedge P   , \\
G_4(P,\sigma_1) &=& 5 [\sigma_1] - \frac{1}{2}( 5U + (4E_1 + 3E_2 + 2E_3 + E_4) - 2\theta) \wedge P,
\eea
where, for now, the overall normalization is chosen to give manifestly integral chiral indices as will be discussed later.
Both flux solutions are not independent on the hypersurface so that we can stick to, say, $G_4(P,\sigma_0)$ for definiteness.



We next address the problem of describing all independent vertical fluxes on the $SU(5) \times \mathbb Z_2$ fibration which exist over a \emph{generic} base ${\cal B}$.
We follow the strategy in \cite{Krause:2012yh}, where the first such classification of vertical gauge fluxes has been undertaken for ($U(1)$ restricted) Tate models with gauge groups $SU(N) (\times U(1))$  for $N=2,3,4,5$ over a generic base ${\cal B}$. See  \cite{Cvetic:2013uta,Bizet:2014uua,Braun:2014xka} for classifications for other types of fibrations.  
To this end we first compute a basis for the vertical $(2,2)$-forms in the ambient space $X_5$ of $X_4$. To simplify the notation we will from now on omit the pull-back symbol `$\pi^{*}$' whenever there is no ambiguity about a divisor coming from the base. Due to relations between the divisors from the Stanley-Reisner ideal SR given in \eqref{eq:SRi_z2} and from homology relations in the fiber ambient space, not all products of divisors are linearly independent. With the help of the computer-algebra system \texttt{Singular} we can take these relations into account and compute a basis for the quotient ring
\begin{equation}\label{eq:quotient_ring}
H^{(*,*)}(X_5) \cong \frac{\mathbb{C}[D_i]}{SR + HOM} \,,
\end{equation}
where $\mathbb{C}[D_i]$ is the formal polynomial ring with all divisors of $X_5$ as variables.\footnote{Strictly speaking this construction only gives the vertical part -- i.e.~linear combinations of products of divisors -- of the ambient space cohomology $H^{(*,*)}(X_5)$, which however suffices for all the computations we perform here. See \cite{Toappear} for a more careful discussion of the vertical cohomology for fibrations over a generic base.} 
The homology relations $HOM$, which can be read off from the top, are encoded in the scaling relations in table \ref{tab:scalings_z2} and take the form
\begin{equation}
\begin{aligned}\label{eq:HOM-relations-z2}
W &= 2 U + 2  \bar{\mathcal{K}} - [b_2] - E_1 - 2 E_2 - 2 E_3 - E_4\,, \\
V &= U + \bar{\mathcal{K}} - [b_2] - E_2 - 2 E_3 - E_4\,,\\
\Theta &= E_0 + E_1 + E_2 + E_3 + E_4\,.
\end{aligned}
\end{equation}
This basis is then used to make an ansatz for the most general flux. The transversality conditions \eqref{eq:bisection_transv_cond} become a set of equations expressed in intersection numbers on the ambient 5-fold, e.g.
\begin{align*}
	\int_{X_4} G_4 \wedge \pi^{-1} D_a \wedge \pi^{-1}D_b = \int_{X_5} [P_{\mathbb Z_2}^{SU(5)}] \wedge G_4 \wedge \pi^{-1}D_a \wedge  \pi^{-1} D_b \, .
\end{align*}

Intersection numbers like these can be reduced to intersection numbers on the base by employing the Stanley-Reisner ideal and the homology relations, thereby eliminating redundancies due to the known homology relations in the fiber ambient space. The Stanley-Reisner ideal trivially sets many intersections to zero. Likewise, due to the fibration structure, any intersection number with more than 3 divisor classes pulled back from the base will vanish. Let $F_i$ denote all divisor classes related to the fibration, both the toric classes $T_i$ associated with the homogeneous coordinates of the original fiber ambient space $\mathbb P_{112}$ and exceptional divisors $E_i$. For $i,j,k$ distinct, and $D_{a,b,c}$ base divisor classes, the non-vanishing intersections (omitting the wedges) are
\begin{equation}\label{eq:basic_intersection_numbers}
\begin{aligned}
\int_{X_5} T_i T_j D_a D_b D_c &= \frac{1}{V(i,j)}\int_{\mathcal B} D_a D_b D_c \, ,\\
\int_{X_5} E_i F_j F_k D_a D_b & =\frac{1}{V(i,j,k)}\int_{\mathcal B} \Theta D_a D_b \, . 
\end{aligned}
\end{equation}
Here $V(i,j)$, ($V(i,j,k)$) is the lattice volume of the cell spanned by the fan vectors $f_i, f_j, (f_k)$. For the top used here, all cell volumes are one, except the one spanned by $f_u$ and $f_v$ corresponding to the divisors $U$ and $V$, which has volume 2. When the $i,j,k$ are non-distinct, we are dealing with a self-intersection of fibral divisor classes. These can be reduced to transversal intersections by using the homology relations in the ambient fiber space. As an example, consider the reduction
\begin{equation}
\begin{aligned}
&\int_{X_5}  W^2 D_a D_b D_c = \int_{X_5} W (2 U + 2  \bar{\mathcal{K}} - [b_2] - E_1 - 2 E_2 - 2 E_3 - E_4) D_a D_b D_c\\
&  = \int_{X_5} W (2 U + 2  \bar{\mathcal{K}} - [b_2]- E_1 - 2 E_2 - 2 E_3)D_a D_b D_c = 2 \int_{\mathcal B} D_a D_b D_c \,.
\end{aligned}
\end{equation}
This way also (self-)intersections of 3,4 or 5 fibral divisor classes may be computed iteratively and reduced to the cases (\ref{eq:basic_intersection_numbers}). \texttt{Singular} automatically applies this method and reduces the transversality conditions to a system of linear combinations of intersection numbers on the base.

As discussed above, if we demand orthogonality with respect to the Cartan generators, i.e.~(\ref{Eicond}), this effectively replaces $\hat U$ by $U$ in the modified transversality condition (\ref{transversality-nsection1}). The solution to all transversality conditions, expressed in a chosen basis, takes the form
\begin{equation}
\begin{aligned}
&G_4  = \\
& z_1 (5 E_1 E_2 + 4E_2^2 + 2 E_3 E_4 + \frac{1}{2} U \Theta + \Theta^2 + (-1,-3,0,1)_iE_i [b_2] + (1,8,0,-2)_i E_i\bar{\mathcal{K}} \\
&\qquad \quad+\frac{1}{2}(-4,-19,-2,3)_i E_i \Theta) \\
+ &z_2 (5 E_1 E_2 + \frac{5}{2}E_2^2 + \bar{\mathcal{K}}\Theta+ (0,-\frac{5}{2},0,0)_iE_i [b_2] + \frac{1}{2}(-4,7,-2,-1)_i E_i\bar{\mathcal{K}} +(0,-5,0,0)_i E_i \Theta) \\
+ &z_3 (5 E_1 E_2 + 2 E_2^2 + E_3 E_4 - U \Theta + [b_2]\Theta+ (0,-3,-1,0)_iE_i [b_2] + (-2,4,0,-1)_i E_i\bar{\mathcal{K}} \\
&\qquad\quad+(0,-4,0,1)_i E_i \Theta)\\
+ & z_4(E_2 E_4 - E_4 \bar{\mathcal{K}}) \,.
\end{aligned}
\end{equation}
However, the last term is a trivial solution on the hypersurface as can be verified by wedging it with the hypersurface class and employing the homology relations. Furthermore, the terms with coefficients $z_2$ and $z_3$ are identical when restricted to the fourfold, again easily seen using the SR-ideal and homology relations. The most general solution for vertical fluxes is thus expressed as
\begin{equation}\label{eq:gen_vertflux_z2}
\begin{aligned}
&G_4  = z_1 G_4^{z_1} + z_2 G_4^{z_2} = \\
&z_1 (5 E_1 E_2 + 4E_2^2 + 2 E_3 E_4 + \frac{1}{2} U \Theta + \Theta^2 + (-1,-3,0,1)E_i [b_2] + (1,8,0,-2)_i E_i\bar{\mathcal{K}} \\
& \qquad \quad+\frac{1}{2}(-4,-19,-2,3)_i E_i \Theta) \\
+\, & z_2 (5 E_1 E_2 + \frac{5}{2}E_2^2 + \bar{\mathcal{K}}\Theta+ (0,-\frac{5}{2},0,0)E_i [b_2] + \frac{1}{2}(-4,7,-2,-1)_i E_i\bar{\mathcal{K}} +(0,-5,0,0)_i E_i \Theta) \,.
\end{aligned}
\end{equation}
Note again that the normalizations for $G_4^{z_1}$ and $G_4^{z_2}$ is chosen to give manifestly integer chiralities.

\subsection{Fluxes from matter surfaces}\label{sec:z2-matter_surfaces}

\noindent So far we have constructed the most general vertical fluxes by systematically implementing the transversality conditions on a basis of $H^{2,2}_{\rm vert}(X_5)$ and pulling these fluxes back to $X_4$.
From a conceptual point of view, the gauge data can be encoded in rational equivalence classes of four-cycles \cite{Bies:2014sra} whose homology class is dual to $G_4$ viewed as an element of $H^{2,2}(X_4)$. The transversality conditions suggest that natural building blocks for the construction of such four-cycles are the matter surfaces. This approach was, for instance, taken in \cite{Borchmann:2013hta} to construct non-Cartan vertical gauge fluxes. In this section we will analyse the 
 matter surfaces associated with states in the antisymmetric and fundamental representations of $SU(5)$ and relate their cohomology classes to the general vertical flux solution found in the previous section. 
 
 As a general remark, recall that the fiber over the ${\bf 10}$-curve in the base  -- see table \ref{fig:matter_curves_poly4} -- splits into a collection of rational curves intersecting like the nodes of the affine Dynkin diagram of $SO(10)$. Suitable combinations of fibral curves are associated with each of the ten entries of the weight vector of the $\mathbf{10}$-representation, and the curves with opposite orientation give rise to the conjugate weights. In the sequel, when we talk about `the $\mathbf{{{\bar{10}}}}$ surface' we have one particular such fibral cycle fibered over the base curve in mind.  Since different weights differ only by combinations of simple roots, different such four-cycles differ by suitable combinations of resolution divisors restricted to the base curve and we will not need to consider all different choices independently. Similar remarks apply to the ${\bf 5}_A$ and ${\bf 5}_B$ representations and their associated matter surfaces.

\subsubsection[The {$\bf{{\bar{10}}}$} Surface]{The {\boldmath{${\bf{{\bar{10}}}}$}} Surface}\label{10barZ2}

\noindent A representative of the matter surface $[\mathcal C_{{\mathbf{{\bar{10}}}}}]$ is given by the complete ambient intersection $(e_0, e_2,b_1)$. By employing the SR-ideal we find that restricting the hypersurface to $(e_0, e_2)$ implies $b_1 = 0$, and hence we can represent the matter surface by $E_0 \wedge E_2$ in the ambient vertical cohomology. This combination is however not orthogonal to the Cartan divisors, and we have to add correction terms to arrive at a valid flux. An ansatz for the correction term of the form $\sum a_i E_i \bar{\mathcal K} + \lambda \bar{\cal K} \Theta$ turns out to be sufficient. This results in the flux
\begin{equation}
\begin{aligned}
G_4({\bf{{\bar{10}}}}) &= E_0 E_2 -\frac{1}{5}\bar{\mathcal{K}} \Theta - \frac{1}{5}(-2,1,-1,-3)_i E_i \bar{\mathcal K}\\
&= -E_1 E_2 - \frac{1}{2}E_2^2 - \frac{1}{5}\bar{\mathcal{K}} \Theta + \frac{1}{2}E_2 [b_2] - \frac{1}{10}(-4,7,-2,-1)_i E_i \bar{\mathcal K} + E_2 \Theta,
\end{aligned}
\end{equation}
where we have rewritten the first line in the chosen vertical basis. Up to a factor of $-5$ the flux agrees exactly with the flux solution with coefficient $z_2$ in \eqref{eq:gen_vertflux_z2}.

\subsubsection{The ${\bf{{\bar{5}}}^A}$ Surface}\label{sec:5A-surface}
\noindent The homology class of the ${\bf{{\bar{5}}}^A}$ matter surface is not straightforwardly given. Over the matter curve $\Theta \cap \{b_1^2 c_{0,4} - b_{0,2} b_1 c_{1,2} + c_{1,2}^2 = 0\}$ the 
rational fiber of the exceptional divisor $E_3$ splits. This can be seen by solving the second polynomial rationally for $c_{0,4}$  and
inserting this together with $e_3=0$ into the hypersurface equation.
 This locally valid approach is enough for computing the weight of the state in the representation, but in order to construct a global flux the homology class of the rationally fibered surface has to be determined. Using \texttt{Singular} we compute the intersection of the hypersurface with the exceptional divisor $E_3$ and the matter curve in the base as the ideal
\begin{equation}
(P_{\mathbb Z_2}^{SU(5)}, e_3, b_1^2 c_{0,4} - b_{0,2} b_1 c_{1,2} + c_{1,2}^2) \, .
\end{equation}
This ideal prime decomposes into two components, corresponding to states in the fundamental and anti-fundamental representations, respectively. The anti-fundamental surface $\mathcal{C}_{\bf{{\bar{5}}}^A}$ is given as the non-transversal intersection
\begin{equation}\label{eq:fiveA_surface}
\begin{aligned}
\mathcal{C}_{\bf{{\bar{5}}}^A} &= (e_3, b_1^2 c_{0,4} - b_{0,2} b_1 c_{1,2} + c_{1,2}^2, e_0^2 e_1 e_4 u^2 c_{1,2} + w b_1, \\
&e_0^2 e_1 e_4 u^2 b_1 c_{0,4} + w b_{0,2} b_1 - w c_{1,2},
e_0^4 e_1^2 e_4^2 u^4 c_{0,4} + e_0^2 e_1 e_4 u^2 w b_{0,2} +w^2) \,.
\end{aligned}
\end{equation}
To make sense of the matter surface as a transversal intersection of three equations in the ambient 5-fold we employ a trick. By prime decomposing the ideal given by the first three polynomials of the above ideal, i.e.~$(e_3, b_1^2 c_{0,4} - b_{0,2} b_1 c_{1,2} + c_{1,2}^2, e_0^2 e_1 e_4 u^2 c_{1,2} + w b_1)$, two irreducible components are revealed. The first one is the matter surface \eqref{eq:fiveA_surface} itself, and the second is the ideal $(e_3, b_1, c_{1,2})$ with multiplicity two. In homology we can `solve' for the matter surface in terms of the two transversal intersections as
\begin{equation}
[\mathcal{C}_{\bf{{\bar{5}}}^A}] = E_3 \wedge 2 [c_{1,2}] \wedge (W + [b_1]) - 2 \cdot E_3 \wedge [b_1] \wedge [c_{1,2}]\,.
\end{equation}
Having obtained the homology class we may construct a transversal flux solution by making an ansatz of correction terms. However, to compare with the previously obtained vertical flux solutions we would like to represent the matter surface as a vertical (2,2)-form in the ambient space which, when restricted to the hypersurface, gives the class $[\mathcal{C}_{\bf{{\bar{5}}}^A}]$. To obtain the solution in this form we make the ansatz
\begin{equation}
[\mathcal{C}_{\bf{{\bar{5}}}^A}] = E_3 \wedge \left(\sum_i a_i D_i \right) \wedge [P_{\mathbb Z_2}^{SU(5)}]
\end{equation}
where the $D_i$ is a basis for the divisors on $X_4$. By expanding both sides in a basis for $H^{3,3}(X_5)$ in \texttt{Singular} we solve for the $a_i$ and obtain that
\begin{align}\label{eq:5B-4cycle-class}
\begin{split}
E_3 \left(\sum_i a_i D_i \right) = &E_3 (E_3 + 2E_4 - [b_2] + 3 \bar{\mathcal{K}} - 3 \Theta) \\
= &\frac{1}{2}E_2^2 + E_3 E_4 + (0,\frac{1}{2}, -1,0)_i E_i [b_2] + (0,-\frac{1}{2},3,\frac{1}{2})_i E_i \bar{\mathcal{K}} + (0,0,-2,0)_i E_i \Theta
\end{split}
\end{align}
restricts to the ${\bf{{\bar{5}}}^A}$ matter surface on the hypersurface. On the righthand side the solution is given in the chosen basis for $H^{2,2}_{\text{\tiny vert}}$. 

\noindent At this point we are ready to construct a well-defined flux by adding a linear combination of terms with at least one factor coming from the base such that the transversality conditions are satisfied. The result is
\begin{equation}
\begin{aligned}
G_4({\bf{{\bar{5}}}^A}) &= [\mathcal{C}_{\bf{{\bar{5}}}^A}]  + \{\text{correction terms}\} \\
&= \frac{1}{2}E_2^2 + E_3 E_4 + \frac{1}{5} [b_2] \Theta -\frac{3}{5} \bar{\mathcal{K}} \Theta + \frac{2}{5}\Theta^2 \\
& \qquad+ \frac{1}{10}(-4,-3,-2,4)_i E_i [b_2] + \frac{1}{10}(12,19,6,-14)_i E_i \bar{\mathcal K} + \frac{1}{5}(-4,-8,-2,4)_i E_i \Theta\\
& = \frac{2}{5}(G_4^{z_1} - G_4^{z_2}).
\end{aligned}
\end{equation}
The last line relates this flux to one combination of vertical fluxes constructed in the previous section.

\subsubsection{The ${\bf{{\bar{5}}}^B}$ Surface}
\noindent By the same technique, we construct a flux from the ${\bf{{\bar{5}}}^B}$ surface. The homology class, obtained by prime decomposition, is
\begin{equation}
[\mathcal{C}_{\bf{{\bar{5}}}^B}] = E_1 \wedge (2 [b_2] + [c_{2,1}]) \wedge (\bar{\mathcal{K}} + U) - 2 E_1 \wedge \bar{\mathcal{K}} \wedge [b_2] .
\end{equation}
By making a suitable ansatz we find that the element
\begin{equation}
E_1(E_1 + 2 E_2 + [b_2] - \Theta)
\end{equation}
in the ambient vertical cohomology reproduces $[\mathcal{C}_{\bf{{\bar{5}}}^B}]$ when restricted to the hypersurface. Using this representative we construct the transversal flux as
\begin{equation}
\begin{aligned}
G_4({\bf{{\bar{5}}}^B}) &= [\mathcal{C}_{\bf{{\bar{5}}}^B}] + \{\text{correction terms}\} \\
&= E_1 E_2 - E_2^2 + E_3 E_4 - 2 U \Theta + \frac{9}{5} [b_2] \Theta - \frac{6}{5}\bar{\mathcal{K}}\Theta  -\frac{2}{5}\Theta^2  \\
& \qquad + \frac{1}{5}(2,-6,-9,-2)_i E_i [b_2] + \frac{1}{5}(-8,-1,6,-2)_i E_i \bar{\mathcal K} + \frac{1}{5}(4,13,2,6)_i E_i \Theta \\
&=  \frac{1}{5}(-2 G_4^{z_1} + 3 G_4^{z_2}),
\end{aligned}
\end{equation}
where the second term is expanded in the chosen basis, and the last line gives the flux as a linear combination of the vertical flux solutions in \eqref{eq:gen_vertflux_z2}. 


\subsection{Chiralities and non-abelian anomalies} \label{sec_Chiralities}

With the explicit flux solutions and also representatives of the homology classes of the matter surfaces at hand, it is straightforward to compute the induced chiralities for all $SU(5)$ representations. The net chirality $\chi$ of a state in representation ${\bf R}$ of $SU(5)$ induced by a flux $G_4$ is given by
\begin{equation}
\chi({\bf R}) = \int_{[\mathcal{C}_{{\bf R}}]} G_4 \,.
\end{equation}
These integrals lift to intersection numbers in the ambient space upon multiplication with the hypersurface class. Using the techniques described above all these intersections are reduced to intersection numbers on the base. The induced chiralities from the three flux solutions described above are, with respect to the general flux combination $ G_4 = a \, G_4(P, \sigma_0) + z_1 G_4^{z_1} + z_2 G_4^{z_2}$,
\begin{equation}
\begin{aligned}
\chi({\bf{{10}}}) &= \left[-a P  + z_1(-2[b_2] + 12\bar{\mathcal K} - 9\Theta ) + z_2(6 \bar{\mathcal K} - 5 \Theta)\right]\bar{\mathcal K} \Theta, \ \\
\chi({\bf{{\bar{5}}}^A}) &= \left[- a P + z_1(-2 [b_2] - 8 \bar{\mathcal K} + \Theta)  - 4z_2 \bar{\mathcal K}\right]([b_2] - 3\bar{\mathcal K} + 2 \Theta) \Theta,\\
\chi({\bf{{\bar{5}}}^B}) &= \left[ a P ([b_2] -4\bar{\mathcal K} + 2\Theta) + z_1(2[b_2]^2 + 3 [b_2]\Theta - 2(6\bar{\mathcal K}^2 - 5\bar{\mathcal K} \Theta + \Theta^2)) \right. \\
&\hphantom{ = [} + \left. z_2 (4 [b_2] - 6 \bar{\mathcal K} + 3 \Theta) \bar{\mathcal K} \right] \Theta,
\end{aligned}
\end{equation}
where we have suppressed integration over the base. It is easily checked that the $SU(5)$ anomaly condition
\begin{equation}
\chi({\bf{{10}}})  = \chi({\bf{{\bar{5}}}^A}) + \chi({\bf{{\bar{5}}}^B}) 
\end{equation}
is satisfied without further restrictions on $a, z_1$ and $z_2$. In fact, this follows directly from the four-cycle class $[\bar{\bf 10}] + [{\bf {\bar{5}^A}}] + [{\bf {\bar{5}^B}}]$: Due to the homology relations (\ref{eq:HOM-relations-z2}) and SR-ideal (\ref{eq:SRi_z2}) this combination is equal to
\begin{align}\label{eq:z2-su5-anomaly-cycle}
	[P_{SU(5)}] \wedge \left\{ 2 [b_2] \wedge (E_1 + E_2) + \bar{\cal K} \wedge (-E_2 + 3E_3 + E_4) + \Theta \wedge ( E_2 - 2E_3 - E_4) - \Theta \wedge [b_2] \right\} \, .
\end{align}
In this form, it is obvious that \textit{any} valid $G_4$ yields zero upon integration over this cycle. In particular, the cancellation of the pure $SU(5)$ anomaly only requires conditions (\ref{transversality-nsection2})  ($G_4$ does not have two legs along the fiber) and (\ref{Eicond}) ($G_4$ does not break gauge symmetry) since the four-cycle class (\ref{eq:z2-su5-anomaly-cycle}) only involves terms of the form $\pi^{-1}D_a \wedge E_i$ and $\pi^{-1}D_a \wedge \pi^{-1}D_b$.
The missing condition (\ref{transversality-nsection1})  will become relevant in the context of the discrete $\mathbb Z_2$ anomaly to be discussed in section \ref{sec_quantisation}.


In addition to the $SU(5)$ charged states, there are localised states with $\mathbb Z_2$ charge $1$ mod $2$ which transform as singlets under $SU(5)$. These states are localised on the curve called $C_2$ in table \ref{table_matterZ2}, which, as we recall, can be described by an ideal generated by 15 non-transversely intersecting elements \cite{Mayrhofer:2014haa}. The $I_2$-fiber over $C_2$ splits into two rational curves $A$ and $B$ with $[A] = [B]$ in homology. Indeed, both curves are exchanged by a global monodromy over $C_2$ provided the intersection of the monodromy locus of the bisection with $C_2$ is non-empty, as is generically the case \cite{Mayrhofer:2014haa} (see \cite {Martucci:2015dxa,Martucci:2015oaa} for a discussion of the implications of the absence of this monodromy point on $C_2$ in non-generic models). 
The states associated with an M2-brane wrapping $A$ and $B$ have the same quantum numbers. In order to count the number of ${\cal N}=1$ chiral multiplets of the 4-dimensional F-theory vacuum with $\mathbb Z_2$ charge $1$, we must therefore add the zero modes from M2-branes wrapping both fibral curves \cite{Mayrhofer:2014laa}. 
One can separately compute the overlap of $G_4$ with the four-cycle ${\cal C}_A$ or ${\cal C}_B$ given by fibering $A$ or $B$ over $C_2$, and e.g.\ the flux $G_4(P, \sigma_0)$ indeed gives a non-zero result for both individual surfaces \cite{Mayrhofer:2014laa}.
However, in total
\bea
\chi({\bf 1}) = \int_{{\cal C}_A} G_4 + \int_{{\cal C}_B} G_4 =0
\eea
by the transversality condition (\ref{transversality-nsection2}) because $A$ and $B$ sum up to the total fiber class. 
This is the geometric manifestation of the statement that an $SU(5)$ singlet carrying only $\mathbb Z_2$ charge does not admit a notion of chirality, of course.

\section{All vertical fluxes on a ${\rm Bl}^1 \mathbb P_{112}[4]$-fibration}\label{sec:u1_geometry}

\noindent The  bisection  $\mathbb P_{112}[4]$-fibration $X_4$ is related, via a conifold transition \cite{Morrison:2014era,Anderson:2014yva,Klevers:2014bqa,Garcia-Etxebarria:2014qua,Mayrhofer:2014haa,Mayrhofer:2014laa}, to the elliptic  ${\rm Bl}^1 \mathbb P_{112}[4]$-fibration  with Mordell-Weil group of rank 1 of \cite{Morrison:2012ei}. 
In general, in a conifold transition between F/M-theory fourfolds conservation of M2-brane charge dynamically relates the 4-form fluxes on both sides \cite{Braun:2011zm,Krause:2012yh,Intriligator:2012ue}. For the specific transition between the $\mathbb P_{112}[4]$-fibration  and the ${\rm Bl}^1 \mathbb P_{112}[4]$-model without extra non-abelian gauge groups,  the $U(1)$ flux and the $\mathbb Z_2$ flux (\ref{eq:z2_flux_sigma1}) have been successfully matched along these lines in \cite{Mayrhofer:2014haa}. In section \ref{sec_Comparison} we will extend this match to the full set of fluxes constructed in the previous section. This will serve as an additional non-trivial check on the consistency of our construction. As a preparation we need to construct, in this section, the complete set of vertical fluxes on the $U(1)$ side of the transition with which we will compare the flux solutions in the bisection model.

Let us briefly recap the properties of the ${\rm Bl}^1 \mathbb P_{112}[4]$-fibration of \cite{Morrison:2012ei}, but including an extra $SU(5)$ factor following \cite{Mayrhofer:2014haa} (see also \cite{Garcia-Etxebarria:2014qua}).  We start from the model \eqref{eq:hypersurface-su5xZ2-model} and by a complex structure deformation set  $c_{4,1} \equiv 0$. This introduces a singularity in codimension 2, which is resolved by a blow-up in the ambient space. The proper transform describing an elliptically fibered fourfold $Y_4$ reads
\begin{equation}\label{eq:hypersurface-su5xu1-model}
\begin{aligned}
P_{U(1)}^{SU(5)}=&\,e_1 e_2 s w^2 + b_{0,2} s^2 u^2 w e_0^2 e_1^2 e_2 e_4 + b_1 s u v w + b_2 v^2 w e_2 e_3^2 e_4\\
& + c_{0,4} u^4 e_0^4 e_1^3 e_2 e_4^2 + c_{1,2} u^3 v e_0^2 e_1 e_4 + c_{2,1} u^2 v^2 e_0 e_3 e_4 + c_{3,1} u v^3 e_0 e_2 e_3^3 e_4^2,
\end{aligned}
\end{equation}
where $s$ is the blow-up coordinate. The divisor class $S: \{s = 0\}$ is the class of an extra rational section, and $U: \{u =0\}$ is the holomorphic zero-section of the elliptic fibration. The structure of the exceptional coordinates $e_i$ is identical to its counterpart in the bisection model because the toric description of $\mathbb P_{112}$ and ${\rm Bl}^1\mathbb P_{112}$ admit the construction of the same top \cite{Bouchard:2003bu}. For the chosen triangulation we obtain the Stanley-Reisner ideal generators
\begin{equation}
\{u w,v s, ve_1, v e_2, w e_0, w e_4, u e_1, u e_2, u e_3, u e_4, s e_2, s e_3, s e_4, e_0 e_3, e_1 e_3, e_1 e_4\} \, .
\end{equation}
The $U(1)$ generator is determined by the Shioda map as
\bea \label{wU1generator}
\tw_{U(1)} = 5(S-U - \bar{\mathcal K} - [b_2]) + 4E_1 + 3E_2 + 2E_3 + E_4.
\eea
The discriminant 
\begin{equation} \label{discrU1}
\Delta \sim \theta^5 [\, b_1^4 b_2 (b_1 c_{3,1} - b_2 c_{2,1})(b_1^2 c_{0,4} - b_{0,2} b_1 c_{1,2} + c_{1,2}^2) + \mathcal{O}(\theta)]
\end{equation}
indicates four matter curves with $SU(5)$ charged matter. In addition there are two singlet curves, only intersecting the $SU(5)$ divisor $\Theta$ in points. The first one is the curve $C_1: (b_2, c_{3,1})$ of conifold singularities which got resolved in the conifold transition. M2-branes wrapping the irreducible fiber components give rise to states of $U(1)$ charge $\pm 10$ (in the normalization (\ref{wU1generator})), called doubly charged states. The second one is a more complicated locus, denoted $C_2$, over which the fiber is of type $I_2$, similarly as in the bisection model. The states localized along this curve have $U(1)$ charge $\pm 5$ and are referred to as singly charged. In table \ref{table_U1spectrum} we summarize the matter spectrum for this model.

\begin{figure}[t]
\vspace{-2cm}
\centering \def\svgwidth{400pt} 
\hspace{1cm}
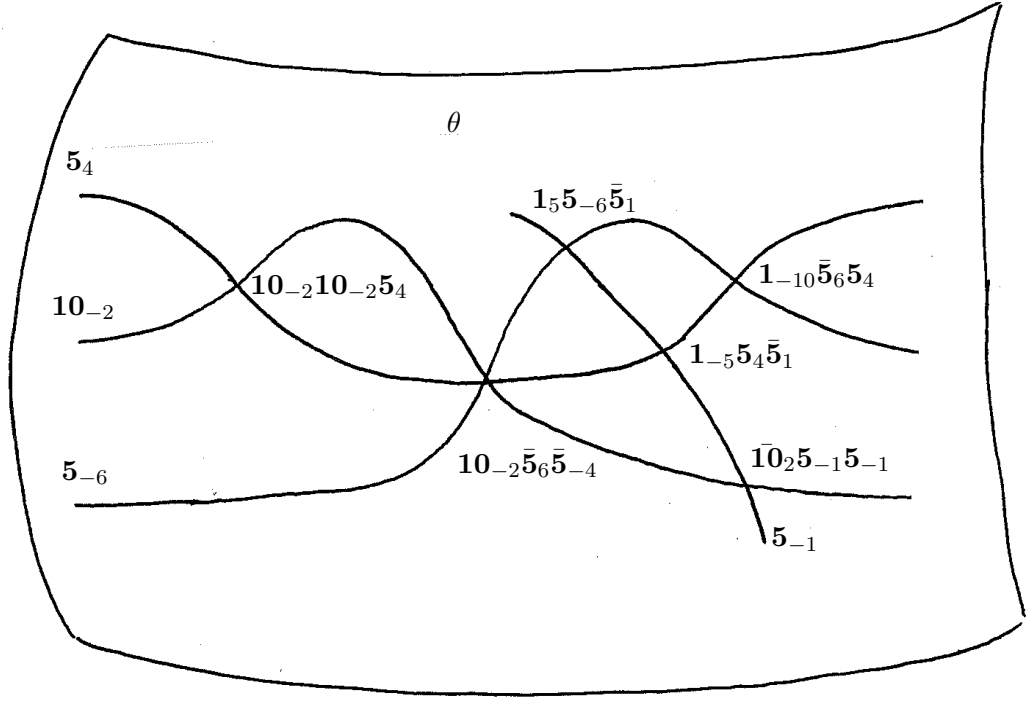 
\vspace{0cm}
\caption{The matter curves in the $SU(5)$ divisor $\{\theta = 0\}$ and the Yukawa couplings involving the $SU(5)$ charged matter in codimension three.}\label{fig:matter_curves_u1model}
\end{figure}

The matter curves intersect at a number of loci, giving rise to 6 different Yukawa couplings involving $SU(5)$ charged fields. These are shown in figure \ref{fig:matter_curves_u1model}. In addition there is one coupling that is localized outside the GUT divisor. This is the coupling ${\bf 1}_{-10} {\bf 1}_5 {\bf 1}_5$ together with its conjugate, and it exists regardless of the $SU(5)$ enhancement. 

\begin{table}[ht]
\begin{center}
\begin{tabular}{ l c } 
locus in base & irrep $SU(5)_{U(1)}$  \\ 
\hline
$\theta \cap b_1$ & ${\bf{10}}_{-2},{\bf{{\bar{10}}}}_2$  \\ 
$\theta \cap b_2$ & ${\bf{5}}_{-6},{\bf{{\bar{5}}}}_6$ \\ 
$\theta \cap \{b_1 c_{3,1} - b_2 c_{2,1}\}$  & ${\bf{5}}_4, \,{\bf{{\bar{5}}}}_{-4}$ \\ 
$\theta \cap \{b_1^2 c_{0,4} - b_{0,2} b_1 c_{1,2} + c_{1,2}^2\}$ & ${\bf{5}}_{-1},{\bf{{\bar{5}}}}_1$  \\ 
$C_1 = b_2 \cap  c_{3,1}$ & $ \bf{1}_{\pm 10} $ \\ 
$C_2$ & $ \bf{1}_{\pm 5} $  \\ 
\end{tabular}
\end{center} 
\caption{Matter curves in the $SU(5) \times U(1)$ model. \label{table_U1spectrum}}
\end{table}

\subsection{All vertical fluxes}


We now construct all vertical flux solutions to the -- in presence of a section -- standard transversality conditions
\bea
\int_{Y_4} G_4 \wedge U \wedge \pi^{-1} D_a     = 0, \quad 
\int_{Y_4} G_4 \wedge D_a \wedge  \pi^{-1} D_b = 0, \quad 
\int_{Y_4} G_4 \wedge E_i \wedge  \pi^{-1} D_a = 0.
\eea 

As always in the presence of a $U(1)$ gauge group, the $U(1)$ generator $\tw_{U(1)}$ in (\ref{wU1generator}) gives rise to a vertical flux solution
\begin{equation}\label{eq:U1-flux}
G_4(F) = \tw_{U(1)} \wedge \pi^{-1}F, 
\end{equation}
which satisfies the transversality conditions for any choice of base divisor class $F$. 

To find more vertical solutions we make a general ansatz, as in the previous section, expressed in a basis for the vertical cohomology of the ambient space $Y_5$. Subjecting this ansatz to the transversality conditions and reducing all terms to intersection numbers in the base we find a family of solutions valid over a generic base ${\cal B}$,
\begin{equation}\label{eq:gen_vertflux_u1}
\begin{aligned}
G_4  =  & G_4(F) +  u_1 G_4^{u_1} +  u_2 G_4^{u_2} +  u_3 G_4^{u_3}   \\
& = \tw_{U(1)} \wedge F    \\
& + u_1 ( - 15 E_1 E_2 +5E_2^2 + 25 E_3 E_4 + (-10,0,-5,10)_i E_i [b_2]\\
& \qquad\qquad\qquad+ (36,37, 18, -16)_i E_i \bar{\mathcal{K}}  + (-20,-25, -10,20)_i E_i \Theta )\\
& + u_2 (-10E_1 E_2 - 5E_2^2 + (0,5,0,0)_i E_i [b_2] + (4,-7,2,1)_i E_i \bar{\mathcal{K}} + (0,10,0,0)_i E_i \Theta) \\
& + u_3 (5E_1 E_2 + 5E_2^2- 5E_3 E_4 +10U \Theta + 10\bar{\mathcal{K}} \Theta + (0,0,5,0)_i E_i [b_2] \\
&\qquad\qquad\qquad+ (-2,1,-6,2)_i E_i \bar{\mathcal{K}} + (-4,-13,-2,-6)_i E_i \Theta). 
\end{aligned}
\end{equation}
The normalization is chosen such as to give manifestly integral chiralities, as presented in the following sections. By restricting the solution to the hypersurface  and expanding it in a basis for $H^{3,3}_{\rm vert}(Y_5)$, it is shown that the three solutions $G_4^{u_i}$ are independent.

\subsection{Fluxes from matter surfaces}
As in the bisection model, it is possible to express all fluxes originating from $SU(5)$ charged matter surfaces in terms of the general vertical flux solution above. In the sequel we derive the map between the two representations of the fluxes. 

\subsubsection{The ${\bf{{\bar{10}}_{2}}}$ surface}

One possible representative for the matter surface $[\mathcal C_{{\bf{{\bar{10}}_{2}}}}]$ is given by the complete ambient intersection $(e_0, e_2,b_1)$, which agrees with the corresponding representation of the ${\bf 10}$-surface considered in the $SU(5) \times \mathbb Z_2$ model. To find the flux associated with this matter surface, we start from an ansatz $E_0 \wedge E_2$ in the ambient space cohomology and add a linear combination of correction terms of the form $U \wedge D_a, S \wedge D_a,  E_i \wedge D_a$ and $D_a \wedge D_b$, for $D_{a,b}$ pullback divisors from the base and solve for the coefficients. Up to the addition of an arbitrary $U(1)$ flux, which we set to zero, the transversality conditions fix the correction terms such that 
\begin{equation}
\begin{aligned}
G_4({\bf{{\bar{10}}_{2}}}) &= E_0 E_2 + \frac{1}{10}(4,-2,2,6)_i E_i \bar{\mathcal K}\\
&= -E_1 E_2 - \frac{1}{2}E_2^2 + \frac{1}{2}E_2 [b_2] + \frac{1}{10}(4,-7,2,1)_i E_i \bar{\mathcal K} + E_2 \Theta \, ,
\end{aligned}
\end{equation}
where we have rewritten the first line in the chosen vertical basis. Up to a scaling factor the flux agrees exactly with the flux solution $G_4^{u_2}$ in \eqref{eq:gen_vertflux_u1}. 

\subsubsection{The ${\bf{{{5}}_{-6}}}$ surface}

A representative of the ${\bf{{{5}}_{-6}}}$ surface is given by the complete intersection of $(e_0, s)$ with the hypersurface. Indeed this implies $b_2 = 0$ and thus reproduces the curve in the base over which the ${\bf{{{5}}_{-6}}}$ matter is localized. 
Repeating verbatim the steps performed for the ${\bf{{\bar{10}}_{2}}}$-flux we arrive at,
\begin{equation}
\begin{aligned}
G_4({\bf{{{5}}_{-6}}}) &= E_0 S - S \Theta + U \Theta + \bar{\mathcal K} \Theta + \frac{1}{5}(4,3,2,1)_i E_i [b_2] - \frac{1}{5}(4,3,2,1)_i E_i \Theta  \\
&=  -E_1 E_2 + U \Theta + \bar{\mathcal K} \Theta + \frac{1}{5}(-1,3,2,1)_i E_i [b_2] + E_1 \bar{\mathcal K} - \frac{1}{5}(4,3,2,1)_i E_i \Theta \\
& = \frac{1}{50}(G_4^{u_1} +6 G_4^{u_2} +  5 G_4^{u_3}). 
\end{aligned}
\end{equation}
In the second line we have used that $E_0 S - S \Theta = - E_1(E_2 - \bar{\mathcal K} + [b_2])$ in the ambient cohomology.


\subsubsection{The ${\bf{{\bar{5}}_{-4}}}$ surface}

The cohomology class of a representative of ${\cal C}_{\bf{{\bar{5}}_{-4}}}$  can be obtained by an ideal decomposition in \texttt{Singular} and is given in the ambient space as
\begin{equation}
 {\mathcal C}_{{\bf{{\bar{5}}_{-4}}}} = E_1 (2 \bar{\mathcal K}^2 + S[b_2] + 2 S \bar{\mathcal K} - S\Theta - \bar{\mathcal K} \Theta ) \, .
\end{equation}
Out of this class a transversal flux may be constructed by adding possible correction terms and solving the transversality conditions. As in the previous chapter we aim at comparing the matter surface to the vertical flux solution. By making the analogous ansatz as in section \ref{sec:5A-surface}, we find that
\begin{equation}\label{eq:5_4_class_ansatz}
{\mathcal C}_{{\bf{{\bar{5}}_{-4}}}}  = E_1 \wedge  (E_1 + 2E_2 + [b_2] -\Theta) \wedge  [P_{SU(5)}] \, .
\end{equation}
The factor of $E_1$ reflects the fact that it is the fiber component of this divisor which splits into weights over the curve. We use this solution to make an ansatz for a well-defined flux in the form of $G_4 = E_1(E_1 + 2E_2 + [b_2] -\Theta) $ +  vertical   correction terms.  As in the previous case, the solution allows for an arbitrary $U(1)$-flux contribution which can be subtracted. There is also a $U(1)$-flux with fixed coefficient appearing and after rewriting the flux in the chosen vertical basis we find the solution
\begin{equation}
\begin{aligned}
G_4({\bf{{\bar{5}}_{-4}}}) &= -E_2^2 + E_3 E_4 - 3 U \Theta - 3 \bar{\mathcal K} \Theta\\
& + \frac{1}{5}(1,-3,-7,-1)_i E_i [b_2] + \frac{1}{5}(-3,-1,6,-2)_i E_i \bar{\mathcal K} + \frac{1}{5}(8,16,4,7)_i E_i \Theta - \frac{1}{5}w_{U(1)} \Theta \\
& = \frac{1}{50}(- G_4^{u_1} - 6 G_4^{u_2} -  15 G_4^{u_3}) -\frac{1}{5} \tw_{U(1)} \Theta.\\
\end{aligned}
\end{equation}
\subsubsection{The ${\bf{{\bar{5}}_{1}}}$ surface}

By the same method we find that
\begin{equation}
{\mathcal C}_{{\bf{{\bar{5}}_{1}}}}  = E_3 \wedge (E_3 + 2E_4 + 3 \bar{\mathcal K} - [b_2] - 2 \Theta)  \wedge  [P_{SU(5)}] \,.
\end{equation}
By adding correction terms we get a well-defined, transversal flux which takes the form
\begin{equation}
\begin{aligned}
G_4({\bf{{\bar{5}}_{1}}}) &= \frac{1}{2}E_2^2 - E_3 E_4 + \frac{1}{10}(-4,-3,-2,4)_i E_i [b_2] + \frac{1}{10}(12,19,6,-7)_i E_i \bar{\mathcal K} + \frac{1}{5}(-4,-8,-2,4)_i E_i \Theta \\
& = \frac{2}{50} (G_4^{u_1} - \frac{3}{2} G_4^{u_2} ).\\
\end{aligned}
\end{equation}

\noindent We conclude with a summary of the full relation between the vertical flux solutions on one side and the matter surface fluxes on the other,
\begin{equation}
\begin{aligned}
G_4({\bf{{\bar{10}}_{2}}})& = \frac{1}{10} G_4^{u_2}, \\
G_4({\bf{{\bar{5}}_{1}}}) &= \frac{2}{50} (G_4^{u_1} - \frac{3}{2} G_4^{u_2} ),\\
G_4({\bf{{\bar{5}}_{-4}}}) &=  \frac{1}{50}(- G_4^{u_1} - 6 G_4^{u_2} -  15 G_4^{u_3}) -\frac{1}{5} \tw_{U(1)} \Theta,\\
G_4({\bf{{{5}}_{-6}}}) &= \frac{1}{50}(G_4^{u_1} +6 G_4^{u_2} +  5 G_4^{u_3})  \, .
\end{aligned}
\end{equation}


\subsection{Chiralities and non-abelian anomalies}


The chiralities induced by the general vertical flux solution $G_4(F) + \sum_i u_i G_4^{u_i}$ are computed as
\begin{equation}
\begin{aligned}
&\chi({\bf{10}}_{-2}) = - 2 [b_1]  F  \Theta + \left[u_1(-20 [b_2] + 42 \bar{\mathcal{K}} - 25 \Theta) + u_2 (-12 \bar{\mathcal{K}} + 10 \Theta ) + u_3(6 \bar{\mathcal{K}}  - 3 \Theta)\right]\bar{\mathcal{K}} \Theta, \\
&\chi({\bf{\bar{5}}}_{1}) =   2  [c_{1,2}] F \Theta + 2\left[ u_1(-10 [b_2] - 14 \bar{\mathcal{K}} + 5 \Theta) + 4 u_2 \bar{\mathcal{K}} + u_3(-2\bar{\mathcal{K}} +\Theta) \right] ([b_2 - 3\bar{\mathcal{K}} + 2 \Theta ])\Theta,  \\ 
&\chi({\bf{\bar{5}}}_{-4}) = [ -4 ([b_2] + [c_{2,1}])  F    +   u_1(10 [b_2]^2 - 16 [b_2]\bar{\mathcal{K}} - 42\bar{\mathcal{K}}^2 + 10 [b_2]\Theta + 61\bar{\mathcal{K}} \Theta - 20\Theta^2 ) \\
&\qquad + 2u_2 \bar{\mathcal{K}}(-2 [b_2] + 6 \bar{\mathcal{K}} - 3\Theta) + u_3(2[b_2]\bar{\mathcal{K}} - 6\bar{\mathcal{K}}^2 + 4[b_2]\Theta + 11\bar{\mathcal{K}}\Theta - 4\Theta^2) ]  \Theta,\\
&\chi({\bf{\bar{5}}}_{6}) = 2\left[  3 F + u_1(5[b_2] - 18 \bar{\mathcal{K}} + 10\Theta ) - 2 u_2  \bar{\mathcal{K}} + u_3( \bar{\mathcal{K}}-3\Theta)\right] [b_2] \Theta\,,
\end{aligned}
\end{equation}
where integration over the base is understood. Consistently, the $SU(5)$ anomaly cancellation condition
\begin{equation}
\chi({\bf{10}}_{-2}) = \chi({\bf{\bar{5}}}_{1}) + \chi({\bf{\bar{5}}}_{-4}) + \chi({\bf{\bar{5}}}_{6})
\end{equation}
holds for all choices of the coefficients $u_i$ and for arbitrary base class $F$.
As in the $\mathbb{Z}_2$ model, we can directly see the $SU(5)$ anomaly cancellation in the geometry because 
\begin{align}
\begin{split}
	&[\bar{\bf 10}_{2}] + [\bar{\bf 5}_{1}] + [\bar{\bf 5}_{-4}] + [\bar{\bf 5}_{6}] = \\
	&[P_{SU(5)}] \wedge \left( 2[b_2] \wedge (E_1 + E_2) + \bar{\cal K}\wedge (-E_2 + 3E_3 + E_4) + \Theta \wedge (-[b_2] + E_2 - E_3 - E_4) \right).
\end{split}
\end{align}
Again this is of the schematic form $E_i \wedge \pi^{-1}D_a + \pi^{-1} D_a \wedge \pi^{-1} D_b$, which yields zero when integrating a valid $G_4$-flux over it.

\section{Comparison via  conifold transition} \label{sec_Comparison}

In this section we compare the flux solutions in the bisection ${\mathbb P}_{112}[4]$-fibration $X_4$ and in the related  elliptic ${\rm Bl}^1{\mathbb P}_{112}[4]$-fibration $Y_4$ upon performing a topological transition between both sides. 
Since the construction of fluxes in F-theory models on elliptic fibrations is well established, as is the topology change in the conifold transition, we will interpret this as another test of our flux construction for the genus-one fibration. In particular, we will construct an explicit map between the flux solutions in both models and show that all fluxes in the bisection model are accounted for by a corresponding flux in the $U(1)$ model upon performing the conifold transition. This map has already been established in \cite{Mayrhofer:2014haa} in absence of additional non-abelian gauge data.

In order to find a map between the general flux solutions, we look for quantities that are preserved under the conifold transition. 
The first such quantity is the total D3-brane charge.
Recall that the number of D3-branes is related to the flux and curvature induced D3-charge  as \cite{Sethi:1996es}
\begin{equation} \label{D3tadpole}
n_{D3} = \frac{\chi(X_4)}{24} - \frac{1}{2}\int_{X_4} G_4 \wedge G_4.
\end{equation}
We are interested in transitions without explicit participation of D3-branes, and for such transitions $n_{D_3}$ must match on both sides of the transition \cite{Gaiotto:2005rp}. We therefore demand that
\begin{equation}\label{eq:delta_d3}
\Delta n_{D3} \equiv n_{D3}|_{\scriptscriptstyle{X_4}} - n_{D3}|_{\scriptscriptstyle{Y_4}} \overset{!}{=} 0 \,.
\end{equation}

The topological transition from $Y_4$ to $X_4$ proceeds by first creating a conifold singularity in the fiber over the curve $C_1 \subset \mathcal{B}$  given in table \ref{table_U1spectrum} and then deforming \cite{Morrison:2014era,Anderson:2014yva,Klevers:2014bqa,Garcia-Etxebarria:2014qua,Mayrhofer:2014haa,Mayrhofer:2014laa}.
The resulting change \cite{Braun:2011zm,Krause:2012yh,Intriligator:2012ue} 
\begin{equation}\label{eq:euler_change}
\Delta \chi = \chi(X_4) - \chi(Y_4) = -3\chi(C_1)
\end{equation}
of  the Euler number allows us to rephrase \eqref{eq:delta_d3} in terms of the flux-induced D3 tadpoles as
\begin{equation}\label{eq:tadpole_change}
\frac{1}{2}\int_{X_4} G_4\wedge G_4 \stackrel{!}{=} -\frac{1}{8} \chi(C_1) + \frac{1}{2} \int_{Y_4} \tilde{G}_4 \wedge\tilde{G}_4 \,.
\end{equation}
Here $G_4$ and $\tilde{G}_4$ denote the fluxes on $X_4$ and $Y_4$, respectively.

The chiral spectra of the two models are topological quantities as well and must be conserved under the transition. 
This applies to the notion of chirality with respect to the unbroken gauge subgroups on both sides of the transition. In the case at hand, this is the non-abelian $SU(5)$ factor.
From the field theory perspective this is clear because  the Higgsing of the $U(1)$ gauge symmetry to a $\mathbb Z_2$ subgroup does not change the $SU(5)$ chiralities of the states. 
However, the number of individual matter curves as such is not equal. By comparing the discriminants (\ref{discrZ2}) for $c_{4,1} \neq 0$ and  (\ref{discrU1}) for $c_{4,1} = 0$, we confirm that the matter curves in the base relate as \cite{Garcia-Etxebarria:2014qua,Mayrhofer:2014haa}
\begin{equation}
\begin{aligned}
X_4  \quad& \quad\quad\quad Y_4 \\
{C}_{\bf{10}} \quad&\leftrightarrow\quad {C}_{\bf{10}_{-2}} \\
{C}_{{\bf{{\bar{5}}}^A}} \quad&\leftrightarrow\quad {C}_{{\bf{{\bar{5}}_{1}}}} \\
{C}_{{\bf{{\bar{5}}}^B}} \quad&\leftrightarrow\quad {C}_{{\bf{{\bar{5}}_{-4}}}} +  \mathcal{C}_{{\bf{{\bar{5}}_{6}}}} \,.
\end{aligned}
\end{equation}
Since the chiral indices are linear in the matter surface classes, we arrive at the following matching condition for the chiral spectra,
\begin{equation}\label{eq:chiral_index_matching}
\begin{aligned}
\chi({\bf{10}}) &\stackrel{!}{=} \chi({\bf{10}_{-2}}), \\
\chi({\bf{{\bar{5}}}^A}) &\stackrel{!}{=} \chi({\bf{{\bar{5}}_{1}}}),\\
\chi({\bf{{\bar{5}}}^B}) &\stackrel{!}{=}  \chi({\bf{{\bar{5}}_{-4}}}) +  \chi({\bf{{\bar{5}}_{6}}}) \, .
\end{aligned}
\end{equation}


To derive the map between the flux solutions
recall first that $C_1 = (b_2, c_{3,1})$ is the doubly charged curve along which the Higgsing is performed. The Euler number of this singlet curve is given by
\begin{equation}
\chi(C_1) = \int_{C_1} c_1(C_1) 
\end{equation}
and with help of the adjunction formula
\begin{equation}
c(C_1) = \frac{c(\mathcal{B})}{1 + [(b_2, c_{3,1})]} \quad \Rightarrow \quad c_1(C_1) = c_1(\mathcal{B}) - [b_2] - [c_{3,1}] = \bar{\mathcal K} -[b_2] - ([\bar{\mathcal K} + [b_2] - \Theta]) = -[c_{4,1}] 
\end{equation}
the Euler number contribution is found as
\begin{equation}
-\frac{1}{8}\chi(C_1) = -\frac{1}{8} \int_\mathcal{B} c_1(C_1) \wedge [c_{3,1}] \wedge [b_2] = \frac{1}{8} \int_\mathcal{B} [b_2]  \wedge [c_{3,1}] \wedge [c_{4,1}] .
\end{equation}


To gain some intuition, let us first consider a flux configuration on $Y_4$ where the flux is simply given by the $U(1)$-flux (\ref{eq:U1-flux}), i.e.~$G_4 = G_4(F)$.
The tadpole contribution on the righthand side of (\ref{eq:tadpole_change}) can then be evaluated as 
\begin{equation}\label{eq:bare_u1_tadpole}
-\frac{1}{8} \chi(C_1) + \frac{1}{2} \int_{Y_4} G_4(F) \wedge G_4(F) = \frac{1}{8} \int_\mathcal{B} [b_2] \wedge [c_{3,1}] \wedge [c_{4,1}] - \int_\mathcal{B} F \wedge F \wedge (\bar{\mathcal{K}} + [b_2] - \frac{2}{5}\Theta).
\end{equation}

From the corresponding transition in \cite{Mayrhofer:2014haa} without $SU(5)$ gauge factor, and also from the general considerations in \cite{Intriligator:2012ue}, we expect that we must allow, possibly amongst other fluxes, for non-vanishing $\mathbb Z_2$-flux $a\, G_4(P,\sigma_0)$ on $X_4$, with a coefficient $a$ to be determined.
Part of the contribution of such $a\, G_4(P,\sigma_0)$ to the lefthand side of  (\ref{eq:tadpole_change})  is given by the square $\frac{1}{2}\int_{X_4} (a \, G_4(P,\sigma_0))^2$ (in addition to cross-terms with the other fluxes). This expression requires in particular the calculation of 
the self-intersection of $[\sigma_0]$. 
The computation proceeds using the normal bundle of $\sigma_0$ embedded in the hypersurface \cite{Braun:2011zm} and closely follows the steps spelled out in \cite{Mayrhofer:2014haa}.
The intersection numbers of $[\sigma_0]$ with the vertical correction term in \eqref{eq:z2_flux_sigma1} are straightforwardly computed in the ambient space, as is the self-intersection of the vertical correction terms. 
After reducing everything to base intersection numbers we obtain
\begin{equation}\label{eq:bare_z2_tadpole}
\frac{1}{2}\int_{X_4} (a G_4(P,\sigma_0))^2 = \frac{25  \, a^2}{4}\int_\mathcal{B}\left( - P \wedge P \wedge (\bar{\mathcal{K}} + [b_2] - \frac{2}{5}\Theta) + 2P \wedge [b_2] \wedge [c_{3,1}] \right).
\end{equation}

Let us first see if it is sufficient to only invoke $a\, G_4(P,\sigma_0)$ in order to reproduce (\ref{eq:bare_u1_tadpole}) on the $\mathbb Z_2$ side, i.e.~whether we can match (\ref{eq:bare_u1_tadpole}) and (\ref{eq:bare_z2_tadpole}).
As seen from \eqref{eq:bare_u1_tadpole}, for a general choice of $F$ the $U(1)$-tadpole has a quadratic term in $\Theta$ from the singlet curve (hidden in the classes $[c_{3,1}]$ and $[c_{4,1}]$), and a linear term in $\Theta$ from the flux contribution. On the other hand, the class $P$ on the $\mathbb Z_2$ side may a priori be dependent or independent of $\Theta$. If it carries no multiple of $\Theta$, then the induced tadpole is only linear in the $SU(5)$ divisor class, which can be excluded. If $P = \ldots + k \Theta$ (which we expect, since $c_{4,1} = \rho \,  \tau$), then the induced tadpole will have a cubic term in $\Theta$, which has to be cancelled in order to match the $U(1)$-tadpole and the singlet curve term. We thus conclude that some other flux has to be turned on in order to satisfy the constraint. In order to see what flux contribution is needed we make the general ansatz 
\begin{equation}\label{eq:gen_flux_z2_model}
G_4 = a \, G_4(P,\sigma_0) + \sum_{i=1}^2 z_i \, G_{4}^{z_i}
\end{equation}
for the flux on the $\mathbb Z_2$-side, with $i$ running over the two solutions \eqref{eq:gen_vertflux_z2}. We furthermore make an ansatz for the class $P = k \, F + \alpha \, [b_2] + \beta \, \bar{\mathcal K} + \gamma \, \Theta$ as a multiple of $F$ plus a correction expanded in the base classes which are \emph{generically} available on any choice of base ${\cal B}$.
The resulting matching equations of induced tadpoles \eqref{eq:tadpole_change} and chiral indices \eqref{eq:chiral_index_matching} are quite lengthy and we do not  display them explicitly here. For  our ansatz above and $u_i = 0$, there is one solution given by
\bea
P = 10F + \frac{1}{2}c_{4,1} , \quad a =\frac{1}{5}, \quad z_1 = -\frac{1}{10}, \quad z_2 = \frac{1}{5}.
\eea
This confirms that it is not enough to turn on only $G_4(P,\sigma_0)$, but that it is also required to allow for  the other vertical fluxes to find a matching configuration. This is in agreement with similar findings in \cite{Krause:2011xj,Krause:2012yh} for 
a transition from an $SU(5) \times U(1)$ elliptic fibration to an $SU(5)$ elliptic fibration.

\noindent Computing the D3-tadpole contributions for a general linear combination of fluxes on both sides of the conifold transition is tedious, but straightforward. We keep the general flux \eqref{eq:gen_flux_z2_model} in the bisection model and since we are searching for the most general solution, we make the ansatz $P = k F + \alpha [b_2] + \beta \bar{\mathcal K} + \gamma \Theta$. In the $U(1)$ model we add the linear combination
\begin{equation} \label{U1ansatzgen}
G_4 = G_4(F) + \sum_{i=1}^3 u_i \, G_4^{u_i}
\end{equation}
of all vertical flux solutions. The reduction of all intersection numbers in \eqref{eq:tadpole_change} and\eqref{eq:chiral_index_matching} to intersection numbers of base divisors results in a system of equations for the coefficients $a, z_i, u_i, k, \alpha, \beta$ and $\gamma$.  The result is that both constraints \eqref{eq:tadpole_change} and \eqref{eq:chiral_index_matching} can be solved by 
\bea
P = 10F + \frac{1}{2}c_{4,1} - 10u_3 \Theta, \quad a = \frac{1}{5}, \quad z_1 = \frac{1}{10}(-1 + 100 u_1), \quad z_2 = \frac{1}{5}(1-65u_1 - 10u_2 + 5u_3) 
\eea
and we further note the $\Theta$-term contribution to the class $P : \{\rho = 0\}$.

It is reassuring that  the possible range $0 \leq P \leq c_{4,1} $ of the divisor class $P = [\rho]$ with $c_{4,1} = \rho \, \tau$  is in beautiful agreement with the observation that fluxes on the $U(1)$ side may obstruct the topological transition provided they induce a purely chiral spectrum of Higgs states \cite{Krause:2012yh,Intriligator:2012ue}. 
The Higgs fields are the charged singlets localised on the curve $C_1$.
The formalism of \cite{Bies:2014sra} suggests that these are counted by the cohomology groups of a line bundle ${\cal L} \otimes K_{C_1}^{1/2}$ with ${\rm deg}({\cal L}) = \int_{C_1} (10 F - 10 u_3 \Theta)$.
This is in agreement with a direct computation of the chiral spectrum of these states, starting from the general flux ansatz (\ref{U1ansatzgen}). 
A necessary condition for the existence of vectorlike pairs of Higgs fields, and thus for the existence of a flat direction for the conifold transition, is that  $ \frac{1}{2} c_1(C_{1}) \leq {\rm deg}({\cal L}) \leq  -\frac{1}{2} c_1(C_{1})$. With $c_1(C_{1})  = - c_{4,1}|_{C_{1}}$ this is in agreement, for the solution $P = 10F + \frac{1}{2}c_{4,1} - 10u_3 \Theta$, precisely with the inequality $0 \leq P \leq c_{4,1}$ -- see the analogous discussion \cite{Mayrhofer:2014haa} in absence of an $SU(5)$ factor. For us, this serves as an additional consistency check of the whole construction.

 \section{Flux quantization and discrete anomalies} \label{sec_quantisation}

All results so far have been independent of the overall normalization of the constructed fluxes and tested only the transversality conditions as such.
The proper normalization becomes crucial for instance when it comes to detecting discrete anomalies such as the ones scrutinized in \cite{Freed:1999vc,Witten:1996md}. In particular, the total number of D3-branes as determined by the tadpole equation (\ref{D3tadpole}) must be integer, and this is guaranteed \cite{Witten:1996md} for a flux satisfying the quantization condition (\ref{quantisation1}).
Furthermore the chiral indices must be integer in a consistent theory and this should follow from the quantization condition as well.
Indeed, as exemplified in previous sections, we can write the homology classes of all matter surfaces ${\cal C}_{\bf R}$ in terms of complete intersections on the hypersurface and so the $[{\cal C}_{\bf R}]$ are integer classes themselves. Hence 
\bea \label{integerchiralities}
\int_{{\cal C}_{\bf R}} \left( G_4 + \frac{1}{2} c_2 (M_4) \right) = \chi({\bf R}) + \frac{1}{2} \int_{{\cal C}_{\bf R}} c_2(M_4) \in \mathbb{Z}
\eea
 if the flux is quantized according to  (\ref{quantisation1}). 
Thus, as stressed in \cite{Krause:2012yh,Collinucci:2012as},  \emph{if} $\frac{1}{2} \int_{{\cal C}_{\bf R}} c_2(M_4)$  is integer by itself for every matter surface, then the quantization condition ensures integrality of the chiral indices. To the best of our knowledge, it has not been proven from first principles in the literature that $c_2(M_4)$ automatically satisfies these constraints in any smooth Calabi-Yau genus-one fibration.
In the sequel will analyze this constraint for the two fibrations $X_4$ and $Y_4$, and relate it to the cancellation of $\mathbb{Z}_2$ anomalies.

\subsection[$c_2(M_4)$ and an arithmetic constraint]{{\boldmath{$c_2(M_4)$}} and an arithmetic constraint}

To compute $c_2(M_4)$ for $M_4$ either the $\mathbb P_{112}[4]$-fibration  $X_4$ or the ${\rm Bl}^1\mathbb P_{112}[4]$-fibration  $Y_4$ we use the standard adjunction formula 
\begin{equation}
c(M_4) = \frac{c(M_5)}{1 + [P]}
\end{equation}
with $P$ the respective hypersurface equation. The answer is expressed in the chosen vertical basis as
\begin{align}
\begin{split}\label{eq:second-chern-class-z2}
c_2(X_4) &= 5U^2-E_1 E_2 + \frac{7}{2}E_2^2-6E_3E_4 + \frac{1}{2}(-4,9,20,4)_i E_i [b_2] + \frac{1}{2}(0,-19,-34,-3)_i E_i \bar{\mathcal{K}}\\
& + (0,-6,4,-5)_i E_i \theta -5U[b_2] +11 U\bar{\mathcal{K}}+  7 U \theta \\
& -6[b_2]\theta -5[b_2]\bar{\mathcal{K}}+ 7\bar{\mathcal{K}} \theta + [b_2]^2 + 5 \bar{\mathcal{K}}^2 + c_2(\mathcal{B}),
\end{split}\\
\begin{split}\label{eq:second-chern-class-u1}
c_2(Y_4) &= -7 U^2 + E_2^2 - E_3 E_4 + (-1,2,5,2)_i E_i [b_2] + (-1,-7,-12,-4)_i E_i \bar{\mathcal{K}}\\
& + (0,-1,4,0)_i E_i \theta  + U[b_2] - U\bar{\mathcal{K}}+  2 U \theta -S[b_2] + 6S \bar{\mathcal{K}} + S \theta\\
& -[b_2]\theta -5[b_2]\bar{\mathcal{K}}+ 2\bar{\mathcal{K}} \theta + [b_2]^2 + 5 \bar{\mathcal{K}}^2 + c_2(\mathcal{B})\,.
\end{split}
\end{align}
Recall that the change in Euler characteristic between the two geometries is given by the Euler number of the doubly charged singlet curve. This provides a cross-check of the Chern classes computed above. The arithmetic genus $\chi_0 = 1+h^{1,0}-h^{2,0}+ ...$ is given by the integral of the Todd class over the fourfold,
\begin{equation}
\chi_0 = \int_{M_4} Td(M_4) = \frac{1}{720}\int_{M_4} 3c_2^2 - c_4 = \frac{1}{720}\left[ \int_{M_4} 3c_2^2  - \chi(M_4) \right] \,.
\end{equation}
For a Calabi-Yau fourfold the arithmetic genus is $\chi_0 = 2$, from which one gets a relation between the squared second Chern class and the Euler characteristic. In particular, for the change in Euler characteristic we have
\begin{equation}
\frac{1}{3}\Delta \chi =   \int_{X_4} c_2(X_4)^2 - \int_{Y_4} c_2(Y_4)^2 \,.
\end{equation}
In the conifold transition we have the relation \eqref{eq:euler_change}, which in terms of the second Chern classes reads
\begin{equation}\label{eq:delta_Euler_char_su5}
\int_{X_4} c_2(X_4)^2 - \int_{Y_4} c_2(Y_4)^2 = -\chi(C_1) = \int_{\mathcal{B}} [b_2] \wedge [c_{3,1}] \wedge [c_{4,1}]\,.
\end{equation}
Given the second Chern classes above it is straightforward to check that \eqref{eq:delta_Euler_char_su5} indeed holds.

Note furthermore that for the quantization condition only $c_2(M_4)$ modulo even forms is relevant. In \cite{Collinucci:2010gz} it was shown that  $c_2({\cal B}) - \bar{\cal K}^2$ is an even class for smooth complex threefolds so that the terms $5 \bar{\mathcal{K}}^2 + c_2(\mathcal{B})$  in $c_2(X_4)$ and $c_2(Y_4)$ can be eliminated mod 2.
In principle the quantization condition can now be checked by demanding that the integral of $G_4 + \frac{1}{2} c_2(M_4)$ over every integer four-cycle be integer. This requires finding an integral basis of $H^4(M_4)$, which we do not attempt here. 

However, we make a curious observation: 
For the elliptic fibration $Y_4$, the integral of $c_2(Y_4)$ over the matter surfaces can be evaluated as 
\bea
\frac{1}{2} \int_{{\cal C}_{{\bf \bar{10}}_2}} c_2(Y_4) &=& \frac{1}{2}  \int_{\cal B} \Theta^2 {\bar{\cal K}} \label{c2overmatterU1a}, \\
\frac{1}{2} \int_{{\cal C}_{{\bf {5}}_{-6}}} c_2(Y_4)   &=&  \frac{1}{2}  \int_{\cal B}( -{\bar {\cal K}} [b_2] \Theta+[b_2]^2 \Theta+ [b_2] \Theta^2 )\label{c2overmatterU1b},\\
\frac{1}{2} \int_{{\cal C}_{{\bf \bar{5}}_{-4}}} c_2(Y_4)   &=&  \frac{1}{2}  \int_{\cal B}  (2 {\bar {\cal K}}^2 \Theta+3 {\bar {\cal K}} [b_2] \Theta+ [b_2]^2 \Theta- {\bar {\cal K}} \Theta^2- [b_2] \Theta^2 ),\label{c2overmatterU1c} \\
\frac{1}{2} \int_{{\cal C}_{{\bf \bar{5}}_1}} c_2(Y_4)   &=&    \int_{\cal B}  (  12 {\bar {\cal K}}^2 \Theta- 10 {\bar {\cal K}} [b_2] \Theta+ 2 [b_2]^2 \Theta-12 {\bar {\cal K}}  \Theta^2+5 [b_2] \Theta^2+3 \Theta^3     ).
\eea 
Note that  the first three expressions are \emph{not} automatically integer. However, in this case also the chiral indices would be non-integer as a result of (\ref{integerchiralities}). Similar expressions can be derived for the singlets.\footnote{A related puzzle was also observed in \cite{Krause:2012yh} for the integral of $\frac{1}{2} c_2$ over the ${\bf 10}_1$-matter surface in the vanilla $SU(5) \times U(1)$ restricted Tate model. Interestingly, existence of a smooth type IIB limit of the latter model implies that this equation is integer, reproducing the known result that the Freed-Witten anomaly cancellation in Type IIB guarantees integer chiralities \cite{Minasian:1997mm,Collinucci:2012as}. }
A similar problem arises in the bisection model $X_4$, where the potentially non-integer pairings are 
\begin{align}\label{eq:z2-c2_over_matter_surfaces}
\begin{split}
	\frac{1}{2} \int_{{\cal C}_{\bf {{\bar{10}}}}} c_2(X_4) &= \frac{1}{2} \int_{\cal B} \! \Theta^2 {\bar {\cal K}} \, ,\\
	\frac{1}{2} \int_{{\cal C}_{\bf {{\bar 5}^A}}} c_2(X_4) &= \int_{\cal B} \! 2 [b_2]^2 \Theta+ 2 \bar{\cal K} [b_2] \Theta - [b_2] \Theta^2 + \bar{\cal K}^2 \Theta - \frac{1}{2} \bar{\cal K} \Theta^2 \, .
\end{split}
\end{align}

Physical consistency therefore requires the expressions (\ref{c2overmatterU1a}), (\ref{c2overmatterU1b}), (\ref{c2overmatterU1c}) (and also the expressions for the singlet surfaces) as well as (\ref{eq:z2-c2_over_matter_surfaces}) to be integer. Note that integrality of (\ref{c2overmatterU1a}) and (\ref{c2overmatterU1b}) of the $U(1)$ model implies integrality of the other expressions including (\ref{eq:z2-c2_over_matter_surfaces}) on the $\mathbb{Z}_2$ side, but integrality of (\ref{eq:z2-c2_over_matter_surfaces}) alone is not enough to guarantee integrality on the $U(1)$ side. We will resolve this puzzle momentarily.

In principle, the above observation could hint at an additional physical constraint such as a previously unnoticed anomaly which could require this. 
A more likely option is that these constraints are automatically satisfied for every smooth Calabi-Yau space $Y_4$ or $X_4$ described as the respective toric tops.
In other words, integrality of the above expressions is most likely a necessary condition for a specific base ${\cal B}$, together with a choice of $\Theta$ and $[b_2]$, to give rise to a well-defined 
Calabi-Yau fibration $Y_4$ or $X_4$.
It would be interesting, but certainly challenging to prove in full generality that in every geometrically consistent fibration $c_2(M_4)$ automatically satisfies these arithmetic properties.


\subsection[Cancellation of ${\bf{\mathbb{Z}_2}}$ anomalies]{Cancellation of {\boldmath{${\bf{\mathbb{Z}_2}}$}} anomalies}

The quantization condition is also crucial in order to investigate possible $\mathbb{Z}_2$ anomalies in the bisection model  and their interplay with the $G_4$-flux.
Due to the charge assignments the possible $\mathbb{Z}_2$ anomalies \cite{Ibanez:1991hv} are given by the chiral index of the ${\bf{{\bar 5}}^A}$ states modulo 2, 
\bea
{\cal A}_{\mathbb Z_2^3 } &=&     \sum_{\bf R} (q_{\bf R}^{\mathbb Z_2})^3 \,\,  {\rm dim}({\bf R})  \,  \chi({\mathbf R}) = \chi({\bf {\bar 5}^A})   \, \, {\rm mod}  \, 2, \\
{\cal A}_{\mathbb Z_2 - SU(5)^2} &=& \sum_{\bf R} q_{\bf R}^{\mathbb Z_2} \, \, c({\bf R}) \, \chi({\mathbf R}) = \chi({\bf \bar 5^A}) \,  \,  {\rm mod}  \, 2, \\
{\cal A}_{\mathbb Z_2 - {\rm grav.} } &=&     \sum_{\bf R} q_{\bf R}^{\mathbb Z_2} \,\,  {\rm dim}({\bf R})  \,  \chi({\mathbf R}) = \chi({\bf \bar 5^A})   \, \, {\rm mod}  \, 2
\eea
with  $c({\bf R})$ the index of the representation.
In general, discrete field theoretic anomalies need not vanish by themselves provided they are cancelled by a suitable discrete version of the Green-Schwarz mechanism \cite{Ibanez:1992ji}. This happens when an anomalous $U(1)$ is Higgsed to a discrete subgroup which is also anomalous. In this case, the anomalous discrete subgroup is not preserved at the non-perturbative level because instantons can violate it.
In our case, however, the $\mathbb Z_2$ symmetry \emph{is} exact at the non-perturbative level. Potential non-perturbative effects would be M2-brane instantons or fluxed M5-instantons. 
The effect of such instantons in the present model has been studied in detail in \cite{Martucci:2015dxa,Martucci:2015oaa}, where it has been shown that they respect the $\mathbb Z_2$ symmetry. This result is in agreement with the general analysis of \cite{BerasaluceGonzalez:2011wy,Camara:2011jg} because the $\mathbb Z_2$ symmetry in question arises from a non-anomalous $U(1)$ via Higgsing \cite{Morrison:2014era,Klevers:2014bqa,Garcia-Etxebarria:2014qua,Mayrhofer:2014haa}. In such cases, string instantons do not break the discrete symmetry further  \cite{BerasaluceGonzalez:2011wy,Camara:2011jg}.
Therefore the mixed $\mathbb Z_2$ symmetries must vanish by themselves.
Consistently, we can adapt the analysis of \cite{Cvetic:2012xn} of the Green-Schwarz mechanism for (mixed) abelian anomalies. The potential Green-Schwarz counter-terms would then be proportional to
\bea
\int_{X_4} G_4 \wedge \hat U \wedge \pi^{-1} D_a.
\eea
As a result of the transversality condition (\ref{transversality-nsection1}) this vanishes identically, confirming once more that the $\mathbb Z_2$ anomalies must vanish by themselves. 

We would like to see the manifestation of this field theoretic argument in the geometry. To this end, we use the homology relations (\ref{eq:HOM-relations-z2}) and the SR-ideal (\ref{eq:SRi_z2}) to rewrite the homology class $[{\cal C}_{\bf{{\bar 5}}^A}]$ as
\begin{align}\label{eq:class_5B-4cycle}
	[P_{SU(5)}] \wedge \left( 2\, E_3 \wedge E_4 - U \wedge \Theta + E_3 \wedge (4 \, \bar{\cal K} - 2 \, [b_2] ) + \Theta \wedge ([b_2] + E_2 - 2\,E_3 + E_4 - \bar{\cal K} ) \right)\, .
\end{align}
In this representation we see that if we impose the transversality conditions (\ref{transversality-nsection1}), (\ref{transversality-nsection2}) and the gauge symmetry condition (\ref{Eicond}) on $G_4$, then we simply have
\begin{align}\label{Z2anomalyrelation}
	\chi( {\bf{{\bar 5}}^A} ) = \int_{X_4} \! G_4 \wedge (2 \, E_3 \wedge E_4) \, .
\end{align}
The question now is whether $\int_{X_4} \! G_4 \wedge E_3 \wedge E_4 \in \mathbb{Z}$ since this would imply that the chirality is even and therefore the discrete $\mathbb Z_2$ anomalies vanish.
For a well-quantized flux satisfying the  quantization condition $G_4 + \frac{1}{2} c_2(X_4) \in H^4(X_4, \mathbb{Z})$, with $c_2(X_4)$ given in (\ref{eq:second-chern-class-z2}), $\mathbb Z_2$ cancellation would follow from $1/2 \int_{X_4} c_2 \wedge E_3 \wedge E_4 \in \mathbb{Z}$, since $E_3 \wedge E_4$ is manifestly integer. A direct calculation reveals that
\begin{align}\label{eq:Z2-anomaly-cancellation}
\begin{split}
	& \int_{X_4} \frac{c_2(X_4)}{2} \wedge E_3 \wedge E_4 = \\
	& \int_{\cal B} \! \Theta \wedge \left(   \frac{1}{2}  (c_2({\cal B}) - \bar{\cal K}^2) - \bar{\cal K}^2 - \Theta^2 \right)- \frac{1}{2} \left([b_2]^2 \, \Theta - 3\,\bar{\cal K} \, [b_2] \, \Theta + 3\, [b_2] \Theta^2 - 5\, \bar{\cal K} \, \Theta^2\right) \, .
\end{split}
\end{align}
While the first term is integer (using the result cited above that $c_2({\cal B}) - \bar{\cal K}^2$ is even), the latter part is not guaranteed to be integer without any further input. 
However, if we assume integrality of all chiral indices in the $U(1)$ model, i.e.~integrality of (\ref{c2overmatterU1a}), (\ref{c2overmatterU1b}) and (\ref{c2overmatterU1c}), then also (\ref{eq:Z2-anomaly-cancellation}) is integral and therefore the discrete $\mathbb Z_2$-anomalies vanish by themselves.
On the other hand, if we impose integrality of chiral indices (\ref{eq:z2-c2_over_matter_surfaces}) as well as the absence of anomalies in the $\mathbb{Z}_2$ model, the arithmetic constraints on the fibration guarantee a consistent (i.e.~integral) chiral spectrum of the $U(1)$ model.



Therefore we see that physical consistency conditions on both the $U(1)$ and the $\mathbb{Z}_2$ model pose \textit{exactly} the same constraints on the geometry. Since the $\mathbb Z_2$ and the $U(1)$ model are related by a conifold transition, it is not surprising that cancellation of the $\mathbb Z_2$ anomalies requires not only integrality of (\ref{eq:z2-c2_over_matter_surfaces}), but of the corresponding expressions in the $U(1)$ model. 
We know that any consistent $\mathbb{Z}_2$ fibration defined by $[b_2]$ and $\Theta$ on the base $\cal B$ originates via Higgsing from a $U(1)$ model over the same base with the same fibration data $[b_2]$ and $\Theta$. Now if the $U(1)$ model is consistent, the chiralities and therefore also (\ref{c2overmatterU1a}), (\ref{c2overmatterU1b}) and (\ref{c2overmatterU1c}) must be integer. These intersection properties of $\cal B$ of course still hold in the $\mathbb{Z}_2$ model and lead to integrality of (\ref{eq:z2-c2_over_matter_surfaces}) as well as the vanishing of the discrete anomaly.
From a field theoretic perspective, cancellation of the discrete anomalies is tied to a consistent embedding of the discrete symmetry into a gauged continuous symmetry at high energies. This underlying gauge symmetry is precisely the $U(1)$ symmetry of the model on $Y_4$ and the relation between consistency of the latter and discrete anomaly cancellation is also expected from this point of view.

Finally, note that the crucial relation (\ref{Z2anomalyrelation}) depends not only on the conditions (\ref{transversality-nsection2}) and  (\ref{Eicond}), as does the proof for cancellation of the non-abelian cubic anomaly, but also on (\ref{transversality-nsection1}), where the bisection appears explicitly. This is our final  consistency check of the transversality conditions.

\section{Conclusions}

In this work we have initiated a systematic investigation of gauge fluxes in F-theory compactifications without section. Such geometries have received considerable attention in the past year \cite{Braun:2014oya,Morrison:2014era,Anderson:2014yva,Klevers:2014bqa,Garcia-Etxebarria:2014qua,Mayrhofer:2014haa,Mayrhofer:2014laa,Cvetic:2015moa} because
they significantly extend the landscape of consistent F-theory vacua beyond the space of elliptic fibrations with a crepant resolution. As one of their attractive features, multi-section fibrations give rise to discrete gauge symmetries upon compactification of M-theory in the F-theory limit.

Our starting point has been a generalization of the known transversality conditions on 4-form fluxes in F-theory models on elliptic fourfolds to compactifications on genus-one fibrations. 
The role of the zero-section in these conditions is replaced by the available multi-section which defines an embedding of a multi-cover of the base into the fourfold. 
We have then put our proposal for the flux consistency conditions to test by constructing all vertical fluxes available for a bisection fibration including an extra non-abelian gauge factor, which for definiteness  we have taken to be $SU(5)$. The total gauge group in F-theory is thus $SU(5) \times \mathbb Z_2$. We have focused on those fluxes  which exist over a generic base ${\cal B}$ without imposing further conditions on the intersection numbers. For a concrete choice of such a base, additional solutions to the transversality conditions may of course arise. We have derived general expressions for the chiral indices of all matter states and confirmed that the transversality conditions automatically imply cancellation of the cubic non-abelian anomalies. As a further test we have dynamically related the constructed fluxes to a basis of vertical fluxes in an F-theory model with gauge group $SU(5) \times U(1)$ which is related to the $SU(5) \times \mathbb Z_2$ model via a conifold transition \cite{Morrison:2014era,Anderson:2014yva,Klevers:2014bqa,Garcia-Etxebarria:2014qua,Mayrhofer:2014haa,Mayrhofer:2014laa}. We have found a perfect match between both sets of fluxes in such a way that a dynamical transition implies a change in the flux quantum numbers without changing the induced M2/D3-brane charge and the chiral indices. This parallels earlier studies performed in \cite{Braun:2011zm,Krause:2012yh,Intriligator:2012ue,Mayrhofer:2014haa}.

A typical challenge in the construction of gauge fluxes is the proper quantization in the sense of \cite{Witten:1996md,Collinucci:2010gz,Collinucci:2012as}. We have shown that a smooth fibration of the type considered must necessarily satisfy a set of arithmetic constraints on certain intersection numbers in the base which guarantee that, independently of the concrete choice of fluxes, all chiral indices are integer. It would be very interesting to prove in full generality that these arithmetic constraints automatically hold on smooth fibrations solely based on geometric arguments. With the help of these relations we have been able to exemplify that the discrete $\mathbb Z_2$ anomalies vanish by themselves. This is in agreement with \cite{BerasaluceGonzalez:2011wy} and the fact that in this geometry non-perturbative effects respect the $\mathbb Z_2$-symmetry \cite{Martucci:2015dxa,Martucci:2015oaa}. 

An obvious next step would be to apply the same reasoning also to genus-one fibrations with higher-degree multi-sections such as the trisection ($\mathbb Z_3$) model studied in \cite{Klevers:2014bqa,Cvetic:2015moa}. From a phenomenological point of view, discrete symmetries are known to be crucial ingredients in MSSM and GUT model building. A systematic search for 3-generation models e.g.\ with gauge group $SU(5) \times \mathbb Z_2$ (with $\mathbb Z_2$ playing the role of R-parity, as exemplified in \cite{Garcia-Etxebarria:2014qua,Mayrhofer:2014haa}) can now be undertaken, along the lines of the global 3-generation examples \cite{Krause:2011xj,Cvetic:2015txa,Toappear} based on elliptic fibrations with other gauge groups.

Finally, recall that  in general, the gauge data associated with the 3-form potential $C_3$ and its 4-form field strength $G_4$ in F/M-theory is encoded \cite{Curio:1998bva,Anderson:2013rka} in the Deligne cohomology group $H^4_{\cal D}(\hat Y, \mathbb Z(2))$.
 A useful parametrization of this rather abstract object can be given in terms of algebraic four-cycles, up to rational equivalence \cite{Bies:2014sra}. When speaking of fluxes, it is typically only the cohomology class that one specifies, but one should keep in mind that this data is sufficient only for the computation of topological quantities such as chiral indices or flux-induced charges. 
A more refined analysis also of the vector-like spectrum, possibly along the lines of \cite{Bies:2014sra} (or, alternatively, \cite{Collinucci:2014taa}), would be desirable and important also for fibrations without section. We hope to address these challenges in the future.


\subsection*{Acknowledgements}

\noindent We thank Martin Bies and Eran Palti for important discussions. The work of LL and OT is supported in part by DFG under Transregio TR-33 `The Dark Universe'  as well as GK 1940 `Particle Physics Beyond the Standard Model' and by Studienstiftung des Deutschen Volkes.
The research of CM is  supported by the Munich Excellence Cluster for Fundamental Physics ``Origin and the Structure of the Universe''.

\appendix
\section{Scalings and divisor classes} \label{app_scaling1}

\noindent In this appendix we present the scaling relations for the coordinates in the two geometries discussed in the paper. For the bisection model described by the hypersurface equation \eqref{eq:hypersurface-su5xZ2-model} the toric coordinates scale as presented in Table \ref{tab:scalings_z2}. For the model with an extra section with hypersurface equation \eqref{eq:hypersurface-su5xu1-model} the scaling relations are collected in Table \ref{tab:scalings_u1}. 
\begin{table}[h]
\centering 
\begin{tabular}{c|c c c c c c c c}
 $$ &$u$ & $v$ & $w$ & $e_0$ & $e_1$ & $e_2$ & $e_3$&$e_4$  \\
\hline 
$\bar{\mathcal K} $ &$\cdot$& $1$ &  $2$ &   $\cdot$ &   $\cdot$ &    $\cdot$ &   $\cdot$ &   $\cdot$\\
$[b_2]$ & $\cdot$& $-1$ &  $-1$ &   $\cdot$ &   $\cdot$ &    $\cdot$ &   $ \cdot$ &   $\cdot$\\
$\theta$ & $\cdot$& $\cdot$ & $\cdot$ & $1$ & $\cdot $ & $\cdot$ & $\cdot$ & $\cdot$\\
\hline
$U$ &	$1$& $1$ &  $2$ &   $\cdot$ & 		$\cdot$ & 	$\cdot$ & 	$\cdot$ &	$\cdot$\\
$E_1$ &	$\cdot$& $\cdot$ & $-1$ &	$-1$ & 	$1$ & 	$\cdot$ &	$\cdot$ &	$\cdot$\\
$E_2$ &	$\cdot$& $-1$ & $-2$ &   	$-1$ & 	$\cdot$ & 	$1$ &   	$\cdot$ &	$\cdot$\\
$E_3$ &	$\cdot$& $-2$ & $-2$ &   	$-1$ & 	$\cdot$ & 	$\cdot$ &	$1$ &	$\cdot$\\
$E_4$ &	$\cdot$& $-1$ & $-1$ &   	$-1$ & 	$\cdot$ & 	$\cdot$ &	$\cdot$ &	$1$\\
\end{tabular}
\caption{\textit{Scaling relations for the toric coordinates in the $\mathbb Z_2$-model.}}
\label{tab:scalings_z2}
\end{table}

\begin{table}[h]
\centering 
\begin{tabular}{c|c c c  c c c c c c}
 $ $ &$u$ & $v$ & $w$ & $s$ & $e_0$ & $e_1$ & $e_2$ & $e_3$&$e_4$  \\
\hline
$\bar{\mathcal K} $ &$\cdot$& $1$ &  $2$& $\cdot$  &   $\cdot$ &   $\cdot$ &    $\cdot$ &   $\cdot$ &   $\cdot$\\
$[b_2]$ & $\cdot$& $-1$ &  $-1$& $\cdot$  &   $\cdot$ &   $\cdot$ &    $\cdot$ &   $ \cdot$ &   $\cdot$\\
$\theta$ & $\cdot$& $\cdot$ & $\cdot$& $\cdot$  & $1$ & $\cdot $ & $\cdot$ & $\cdot$ & $\cdot$\\
\hline
$U$ &	$1$& $1$ &  $2$& $\cdot$  &   $\cdot$ & 		$\cdot$ & 	$\cdot$ & 	$\cdot$ &	$\cdot$\\
$S$ &	$\cdot$& $1$ &  $1$& $1$  &   $\cdot$ & 		$\cdot$ & 	$\cdot$ & 	$\cdot$ &	$\cdot$\\
$E_1$ &	$\cdot$& $\cdot$ & $-1$& $\cdot$  &	$-1$ & 	$1$ & 	$\cdot$ &	$\cdot$ &	$\cdot$\\
$E_2$ &	$\cdot$& $-1$ & $-2$& $\cdot$  &   	$-1$ & 	$\cdot$ & 	$1$ &   	$\cdot$ &	$\cdot$\\
$E_3$ &	$\cdot$& $-2$ & $-2$& $\cdot$  &   	$-1$ & 	$\cdot$ & 	$\cdot$ &	$1$ &	$\cdot$\\
$E_4$ &	$\cdot$& $-1$ & $-1$& $\cdot$  &   	$-1$ & 	$\cdot$ & 	$\cdot$ &	$\cdot$ &	$1$\\
\end{tabular}
\caption{\textit{Scaling relations for the toric coordinates in the $U(1)$-model.}}
\label{tab:scalings_u1}
\end{table}
\newpage
\bibliography{papers}  
\bibliographystyle{custom1}

\end{document}

%% file: mattercurvesF4.pdf_tex
\begingroup%
  \makeatletter%
  \providecommand\color[2][]{%
    \errmessage{(Inkscape) Color is used for the text in Inkscape, but the package 'color.sty' is not loaded}%
    \renewcommand\color[2][]{}%
  }%
  \providecommand\transparent[1]{%
    \errmessage{(Inkscape) Transparency is used (non-zero) for the text in Inkscape, but the package 'transparent.sty' is not loaded}%
    \renewcommand\transparent[1]{}%
  }%
  \providecommand\rotatebox[2]{#2}%
  \ifx\svgwidth\undefined%
    \setlength{\unitlength}{469.18754883bp}%
    \ifx\svgscale\undefined%
      \relax%
    \else%
      \setlength{\unitlength}{\unitlength * \real{\svgscale}}%
    \fi%
  \else%
    \setlength{\unitlength}{\svgwidth}%
  \fi%
  \global\let\svgwidth\undefined%
  \global\let\svgscale\undefined%
  \makeatother%
  \begin{picture}(1,0.58555134)%
    \put(0,0){\includegraphics[width=\unitlength]{mattercurvesF4.pdf}}%
    \put(0.49201675,0.46912136){\color[rgb]{0,0,0}\makebox(0,0)[lb]{\smash{$\theta$}}}%
    \put(0.09508666,0.12826193){\color[rgb]{0,0,0}\makebox(0,0)[lb]{\smash{${\bf{10}}^{(0)}$}}}%
    \put(0.20133084,0.46839303){\color[rgb]{0,0,0}\makebox(0,0)[lb]{\smash{${\bf{5}}_A^{(1)}$}}}%
    \put(0.73896675,0.41710062){\color[rgb]{0,0,0}\makebox(0,0)[lb]{\smash{${\bf{5}}_B^{(0)}$}}}%
    \put(0.22029026,0.31000283){\color[rgb]{0,0,0}\makebox(0,0)[lb]{\smash{$\bar{\bf{10}}^{(0)}\,{\bf{5}}_A^{(1)}\,{\bf{5}}_A^{(1)}$}}}%
    \put(0.37750665,0.13530468){\color[rgb]{0,0,0}\makebox(0,0)[lb]{\smash{${\bf{1}}^{(1)}\,{\bf{5}}_A^{(1)}\,{\bar{\bf{5}}}_B^{(0)}$}}}%
    \put(0.53211582,0.27301893){\color[rgb]{0,0,0}\makebox(0,0)[lb]{\smash{$\bar{\bf{10}}^{(0)}\,{\bf{5}}_B^{(0)}\,{\bf{5}}_B^{(0)}$}}}%
    \put(0.69072633,0.18898005){\color[rgb]{0,0,0}\makebox(0,0)[lb]{\smash{${\bf{10}}^{(0)}\, {\bf{10}}^{(0)}\, {\bf{5}}_B^{(0)}$}}}%
  \end{picture}%
\endgroup%

%% file: mattercurvesU1model.pdf_tex
\begingroup%
  \makeatletter%
  \providecommand\color[2][]{%
    \errmessage{(Inkscape) Color is used for the text in Inkscape, but the package 'color.sty' is not loaded}%
    \renewcommand\color[2][]{}%
  }%
  \providecommand\transparent[1]{%
    \errmessage{(Inkscape) Transparency is used (non-zero) for the text in Inkscape, but the package 'transparent.sty' is not loaded}%
    \renewcommand\transparent[1]{}%
  }%
  \providecommand\rotatebox[2]{#2}%
  \ifx\svgwidth\undefined%
    \setlength{\unitlength}{419.36770955bp}%
    \ifx\svgscale\undefined%
      \relax%
    \else%
      \setlength{\unitlength}{\unitlength * \real{\svgscale}}%
    \fi%
  \else%
    \setlength{\unitlength}{\svgwidth}%
  \fi%
  \global\let\svgwidth\undefined%
  \global\let\svgscale\undefined%
  \makeatother%
  \begin{picture}(1,0.66030741)%
    \put(0,0){\includegraphics[width=\unitlength]{mattercurvesU1model.pdf}}%
    \put(0.43239569,0.53796504){\color[rgb]{0,0,0}\makebox(0,0)[lb]{\smash{$\theta$}}}%
    \put(0.07555751,0.50205883){\color[rgb]{0,0,0}\makebox(0,0)[lb]{\smash{${\bf{5}}_4$}}}%
    \put(0.06231959,0.36543203){\color[rgb]{0,0,0}\makebox(0,0)[lb]{\smash{${\bf{10}}_{-2}$}}}%
    \put(0.0719033,0.21012325){\color[rgb]{0,0,0}\makebox(0,0)[lb]{\smash{${\bf{5}}_{-6}$}}}%
    \put(0.24853219,0.38552023){\color[rgb]{0,0,0}\makebox(0,0)[lb]{\smash{${\bf{10}}_{-2}{\bf{10}}_{-2}{\bf{5}}_4$}}}%
    \put(0.44210601,0.21682945){\color[rgb]{0,0,0}\makebox(0,0)[lb]{\smash{${\bf 10}_{-2} \bar{{\bf 5}}_6 \bar{{\bf 5}}_{-4}$}}}%
    \put(0.51194896,0.46672754){\color[rgb]{0,0,0}\makebox(0,0)[lb]{\smash{${\bf 1}_{5}{\bf 5}_{-6}\bar{{\bf 5}}_1$}}}%
    \put(0.65890863,0.32003865){\color[rgb]{0,0,0}\makebox(0,0)[lb]{\smash{${\bf 1}_{-5}{\bf 5}_4\bar{{\bf 5}}_{1}$}}}%
    \put(0.71617984,0.2222794){\color[rgb]{0,0,0}\makebox(0,0)[lb]{\smash{$\bar{{\bf 10}}_2 {\bf 5}_{-1}{\bf 5}_{-1}$}}}%
    \put(0.73605854,0.14737198){\color[rgb]{0,0,0}\makebox(0,0)[lb]{\smash{${\bf 5}_{-1}$}}}%
    \put(0.72375177,0.39587257){\color[rgb]{0,0,0}\makebox(0,0)[lb]{\smash{${\bf 1}_{-10} \bar{{\bf 5}}_6 {\bf 5}_4$}}}%
  \end{picture}%
\endgroup%